\documentclass[a4paper,11pt]{article}
\usepackage{jinstpub}
\usepackage{lineno}
\usepackage{booktabs}
\usepackage{multirow}
\usepackage{siunitx}
\usepackage{xcolor}
\definecolor{revisiongreen}{RGB}{0,128,0}
\newcommand{\revgreen}[1]{\textcolor{revisiongreen}{#1}}
\usepackage{subcaption}
\usepackage{booktabs}
\usepackage{longtable}
\usepackage{array}
\usepackage{pdflscape}
\usepackage{multirow}
\usepackage{ragged2e}
\usepackage[normalem]{ulem}

\title{\boldmath Expected performance of a water Cherenkov detector for reactor antineutrino--electron scattering}

\author[a]{Sunhong Kim,}
\author[b]{Sunwoo Gwon,}
\author[c]{Youngju Ko,}
\author[a]{Hokyeong Nam,}
\author[a]{Juseong Park,}
\author[a]{and Yujin Park,}
\author[a,1]{Kim Siyeon,\note{Corresponding author.}}

\affiliation[a]{Department of Physics, Chung-Ang University,
Seoul, Republic of Korea}
\affiliation[b]{Department of Physics, Chonnam National University,
Gwangju, Republic of Korea}
\affiliation[c]{Department of Physics, Jeju National University,
Jeju, Republic of Korea}

\emailAdd{goodgoose@cau.ac.kr}
\emailAdd{tnsdn302@gmail.com}
\emailAdd{yjkophys@jejunu.ac.kr}
\emailAdd{hokyeong97@cau.ac.kr}
\emailAdd{juseongpark0921@gmail.com}
\emailAdd{yjp143680@cau.ac.kr}
\emailAdd{siyeon@cau.ac.kr}

\abstract{
We evaluate the expected performance of a \SI{170}{t} water Cherenkov
detector for reactor antineutrino--electron elastic scattering at the RENO
near-detector site. The reactor-spectrum normalization is extracted from the
reconstructed recoil-electron directional distribution using a constrained
profile-likelihood fit for an exposure of 365.25 days. For the minimum prompt-hit multiplicity
$N_{\mathrm{prompt}}\geq8$ scenario, which neglects PMT radioactivity, the
total uncertainty on the reactor-spectrum normalization is
${}^{+1.08\%}_{-1.07\%}$.For the more restrictive
$N_{\mathrm{prompt}}\geq20$ scenario, which includes the simulated
PMT-radioactivity components, the total uncertainty is
${}^{+1.49\%}_{-1.47\%}$. These results demonstrate that directional
information from a water Cherenkov detector can provide percent-level
sensitivity to the reactor elastic-scattering normalization.
}


\begin{document}
\maketitle
\flushbottom


\section{Introduction}
\label{sec:introduction}

Nuclear reactors provide intense and continuous sources of electron
antineutrinos with energies of a few MeV. Most reactor-antineutrino
experiments detect these particles through inverse beta decay (IBD),
\begin{equation}
\bar{\nu}_{e}+p\rightarrow e^{+}+n,
\end{equation}
which offers a comparatively large interaction cross section and a distinctive
prompt--delayed coincidence signature. These features have enabled precise
measurements of reactor-antineutrino disappearance, flux, and energy spectra,
including those performed by KamLAND~\cite{KamLAND2003},
Daya Bay~\cite{DayaBay2012}, the RENO experiment at the Hanbit Nuclear Power
Plant~\cite{RENOFlux2021}, and NEOS~\cite{NEOS2017}. The IBD
channel nevertheless has an incident antineutrino-energy threshold of
approximately \SI{1.806}{MeV}, intrinsically restricting IBD-based flux and
spectral measurements to the portion of the reactor-antineutrino spectrum
above this threshold. In addition, the positron direction at reactor energies
is only weakly correlated with the incident antineutrino
direction~\cite{VogelBeacom1999}. Consequently, although IBD has proven highly
effective for rate and spectral measurements, it offers limited sensitivity
to the incident antineutrino direction.

Elastic scattering of electron antineutrinos on atomic electrons,
\begin{equation}
    \bar{\nu}_{e}+e^{-}\rightarrow\bar{\nu}_{e}+e^{-},
\end{equation}
provides a complementary detection channel.  Unlike IBD, this process has no
reaction threshold associated with the production of a neutron and positron;
the practical threshold is instead determined by the recoil-electron energy
required for detection and event reconstruction. The interaction is a leptonic electroweak process whose cross section can be calculated precisely within the Standard Model~\cite{VogelEngel1989,TomalakHill2020}.  More
importantly for the present study, the recoil electron is predominantly
emitted in the forward direction relative to the incident antineutrino.

A water Cherenkov detector is a natural instrument for exploiting this
forward-scattering signature.  Water provides a large number of target
electrons at relatively low material cost, while prompt Cherenkov photons
encode both the interaction vertex and the recoil-electron direction.  The
principal challenge is that the antineutrino--electron elastic-scattering (ES)
cross section is substantially smaller than the IBD cross section, and the
single-electron final state does not provide a delayed-coincidence signature.
Radioactivity in the photosensors and detector materials, dissolved radon,
and cosmogenic beta-emitting isotopes can therefore produce large samples of
electron-like background events. At a short-baseline reactor site,
IBD interactions can provide an additional reactor-correlated background when
the prompt positron is retained but the accompanying neutron is not tagged.
Directional information is therefore particularly important in this regime,
as the reactor ES signal is correlated with the known reactor-core directions,
whereas radioactive and cosmogenic backgrounds are not expected to
exhibit such a correlation and the IBD positron direction is only weakly
correlated with the incident antineutrino direction.

The feasibility of reconstructing reactor-antineutrino directions through ES
in large gadolinium-loaded water Cherenkov detectors has previously been
studied for long-baseline reactor-monitoring applications~\cite{Hellfeld2017}.
Those studies identified water-borne radon, cosmogenic isotopes, and external
or photosensor radioactivity as important experimental limitations. The
background model adopted in the present study includes these principal
components. Unlike Ref.~\cite{Hellfeld2017},we additionally simulate IBD events as a background component. The IBD event response is explicitly modeled, whereas the neutron-tagging efficiency is treated parametrically using assumed values rather than being simulated directly. We evaluate the expected performance of a \SI{170}{t}
pure-water Cherenkov detector located at the RENO near-detector site. The short
baselines to the six Hanbit reactor cores provide a high ES interaction rate,
allowing us to assess whether the total reactor ES normalization can be
constrained through a directional fit to a selected sample dominated by
physical backgrounds.

The remainder of this paper is organized as follows. Section 2 describes the reference detector configuration, the Hanbit reactor geometry and expected $\bar{\nu_e}$-electron elastic-scattering rate, and the principal physical background sources considered in the study. Section 3 summarizes the Monte Carlo simulation of the detector response, signal events, and radioactive and cosmogenic backgrounds. Section 4 presents the event reconstruction and selection procedures, including prompt-hit selection, vertex reconstruction, directional reconstruction, and the resulting reconstruction performance. Section 5 describes the directional ES analysis, the profile-likelihood framework and nuisance-parameter treatment, the extraction of the reactor-spectrum normalization, the projected event yields and uncertainties, the impact of residual IBD background under
different assumed neutron-tagging efficiencies, the idealized truth-level energy selections, and the statistical validation with pseudo-experiments. Finally, Section 6 summarizes the main results and discusses the principal limitations and requirements for a realistic experimental implementation.

\section{Detector concept and experimental conditions}
\label{sec:detector_conditions}

\subsection{Reference detector geometry}
\label{subsec:detector_geometry}

The reference detector considered in this study is modeled as a right
circular cylinder filled with ultra-pure water.  The cylindrical target
has a diameter and height of \SI{6}{m}, corresponding to a total water
mass of approximately \SI{170}{t}.

A total of 9,480 4-inch photomultiplier tubes (PMTs) are distributed
over the inner surface of the target.  Each PMT is represented by a simplified cylindrical geometry with an outer diameter of
\SI{102}{mm} and an active photocathode diameter of \SI{95}{mm}.  The PMT size and number were chosen to provide approximately $10^{4}$
readout channels and a photocathode coverage of about 39.6\%.
The simplified PMT geometry does not include the detailed physical curvature
of the photocathode.

The top and bottom surfaces each contain 1,545 PMTs with a pitch of
\SI{135}{mm}.  The remaining 6,390 PMTs are arranged on the cylindrical
side surface in a $142\times45$ layout, with pitches of approximately
\SI{129}{mm} azimuthally and \SI{133}{mm} axially.  For this arrangement,
the photocathode coverage is defined as
\begin{equation}
    f_{\mathrm{cov}}
    =
    \frac{
        N_{\mathrm{PMT}} \pi r_{\mathrm{pc}}^{2}
    }{
        2\pi R H + 2\pi R^{2}
    },
    \label{eq:photocathode_coverage}
\end{equation}
where $N_{\mathrm{PMT}}=9480$ is the total number of PMTs,
$r_{\mathrm{pc}}=\SI{47.5}{mm}$ is the photocathode radius,
and $R=\SI{3}{m}$ and $H=\SI{6}{m}$ are the radius and height of the
cylindrical target, respectively. These parameters correspond to a
photocathode coverage of 39.6\%.

\begin{figure}[t]
    \centering

    \begin{subfigure}[t]{0.49\textwidth}
        \centering
        \includegraphics[width=\linewidth]
        {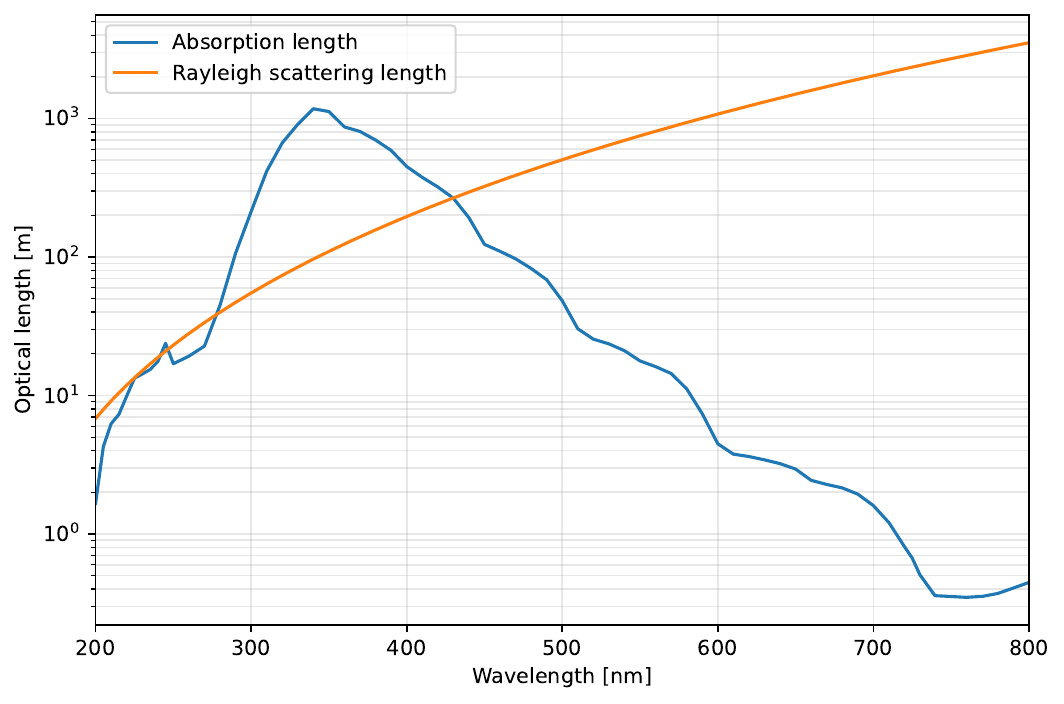}
        \caption{
            Absorption and Rayleigh scattering lengths of water.
        }
        \label{fig:water_optical_lengths}
    \end{subfigure}
    \hfill
    \begin{subfigure}[t]{0.49\textwidth}
        \centering
        \includegraphics[width=\linewidth]
        {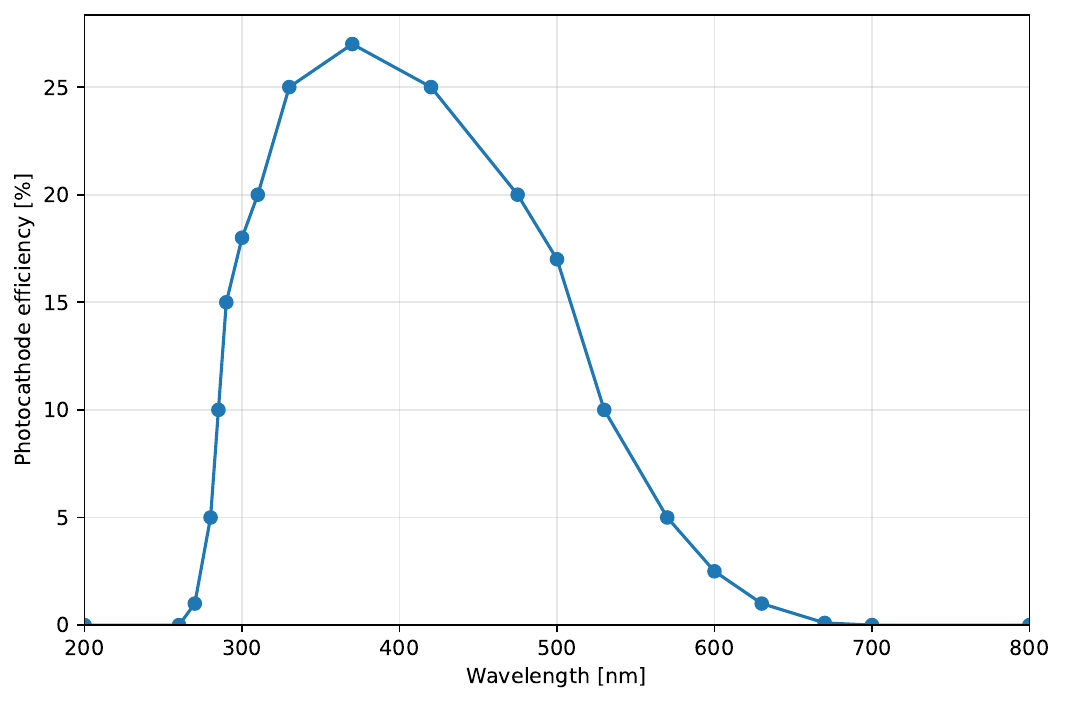}
        \caption{
            Photocathode efficiency adopted from the default
            RAT-PAC optical response.
        }
        \label{fig:photocathode_efficiency}
    \end{subfigure}

    \caption{
        Wavelength-dependent optical properties used in the detector
        simulation.  The water absorption length is taken from
        ref.~\cite{FewellTrojan2019}, while the Rayleigh scattering
        length is calculated according to ref.~\cite{ZhangHu2021}.
        The photocathode efficiency is taken from the default RAT-PAC
        optical response~\cite{RatpacTwo2023}.
    }
    \label{fig:optical_properties}
\end{figure}

Optical-photon propagation in water is simulated using
wavelength-dependent refractive index, absorption length, and Rayleigh
scattering length. The refractive index follows the IAPWS formulation
for water at $T=\SI{295}{K}$ and a density of
\SI{1000}{kg.m^{-3}}~\cite{IAPWS1997}. The absorption length is taken
from ref.~\cite{FewellTrojan2019}, while the Rayleigh scattering length
is calculated according to ref.~\cite{ZhangHu2021}. The photocathode
efficiency is adopted from the default RAT-PAC optical
response~\cite{RatpacTwo2023}, and a single-photoelectron timing
resolution of \SI{1}{ns} is assumed. The wavelength-dependent absorption
and Rayleigh-scattering lengths and the photocathode response are shown in
figure~\ref{fig:optical_properties}. A more detailed numerical tabulation
of the water refractive index, absorption length, and Rayleigh-scattering
length is provided in
appendix~\ref{app:water_optical_properties}. The principal parameters of the reference detector configuration are
summarized in table~\ref{tab:detector_parameters}.

\begin{table}[t]
    \centering
    \caption{Principal parameters of the reference detector configuration.}
    \label{tab:detector_parameters}
    \begin{tabular}{ll}
        \toprule
        Parameter & Reference value \\
        \midrule
        Target geometry
            & Cylinder, \SI{6}{m} diameter
              $\times$ \SI{6}{m} height \\
        Target medium and mass
            & Ultra-pure water, approximately \SI{170}{t} \\
        Photosensor model
            & Reference \SI{4}{in} PMT model \\
        PMT dimensions
            & \SI{102}{mm} outer diameter
              (\SI{95}{mm} active photocathode) \\
        PMT arrangement
            & 9,480 total: 1,545 top, 1,545 bottom, and 6,390 side \\
        Photocathode coverage
            & 39.6\% \\
        Photocathode efficiency
            & Wavelength dependent; 27\% peak at \SI{370}{nm} \\
        Single-photoelectron timing resolution
            & \SI{1}{ns} \\
        \bottomrule
    \end{tabular}
\end{table}

The vessel structure, mechanical support, calibration hardware,
water-circulation system, veto instrumentation, and detailed electronics
design are outside the scope of the present performance projection.
The detector configuration defined above, which represents a reference simulation geometry rather than an optimized engineering design, is used throughout the signal
and background calculations presented in the following subsections.

\subsection{Hanbit reactor configuration and expected signal rate}
\label{subsec:reactor_configuration}

\begin{figure}[t]
    \centering
    \includegraphics[width=0.78\textwidth]{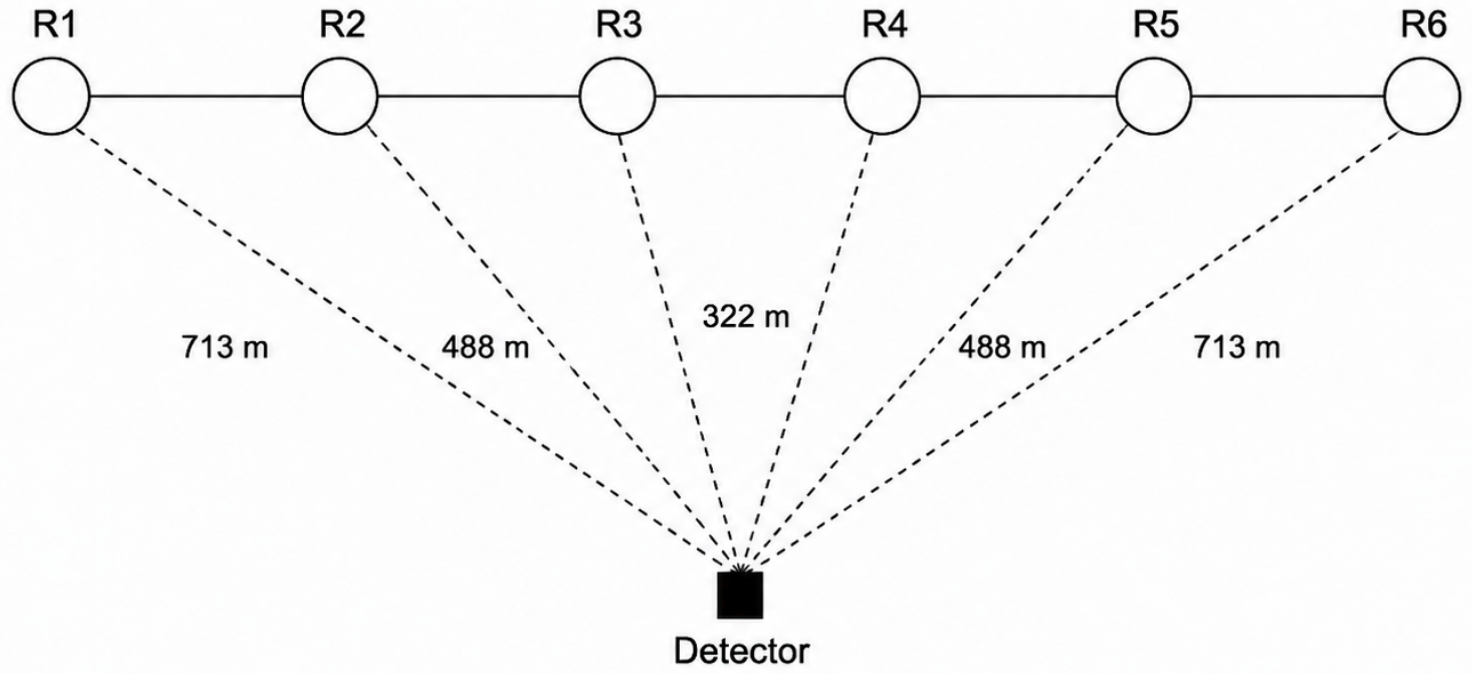}
    \caption{
        Schematic, not-to-scale layout of the six Hanbit reactor cores
        and the detector assumed to be installed at the RENO
        near-detector location.
    }
    \label{fig:hanbit_layout}
\end{figure}

The detector is assumed to be installed at the same location as the
RENO near detector at the Hanbit Nuclear Power Plant.  The relative
layout of the detector and the six reactor cores is illustrated in
figure~\ref{fig:hanbit_layout}.  At this location, cores 3 and 4 are
each at a baseline of approximately \SI{322}{m}, cores 2 and 5 are
located at approximately \SI{488}{m}, and cores 1 and 6 are located at
approximately \SI{713}{m}~\cite{RENOFlux2021}.  Each of the six cores
is assumed to operate continuously at a nominal thermal power of
\SI{2.8}{GW_{th}}. 
Time-dependent variations of the individual reactor powers and fuel
compositions are not modeled; the nominal configuration is used as a fixed
reference throughout this sensitivity study.

The corresponding fission rate of each core is estimated from its
nominal thermal power as
\begin{equation}
    \dot{N}_{f,k}
    =
    \frac{P_{\mathrm{th},k}}{\langle E_f\rangle},
    \label{eq:fission_rate}
\end{equation}
where an average energy release of
$\langle E_f\rangle=\SI{200}{MeV}$ per fission is assumed.  For a
thermal power of \SI{2.8}{GW_{th}}, this gives
$\dot{N}_{f,k}=8.7\times10^{19}\,\mathrm{s^{-1}}$ per core.

The number of target electrons in the \SI{170}{t} water target,
$N_e$, is calculated as
\begin{equation}
    N_e
    =
    Z_{\mathrm{H_2O}}
    \frac{m_{\mathrm{target}}}
         {M_{\mathrm{H_2O}}}
    N_{\mathrm A}
    =
    5.69\times10^{31},
    \label{eq:number_target_electrons}
\end{equation}
where $Z_{\mathrm{H_2O}}=10$ is the number of electrons per water
molecule, $m_{\mathrm{target}}=\SI{170}{t}$ is the water-target mass,
$M_{\mathrm{H_2O}}=\SI{18.015}{g.mol^{-1}}$ is the molar mass of
water, and $N_{\mathrm A}$ is the Avogadro constant.

\begin{figure}[t]
    \centering

    \begin{subfigure}[t]{0.49\textwidth}
        \centering
        \includegraphics[width=\linewidth]
        {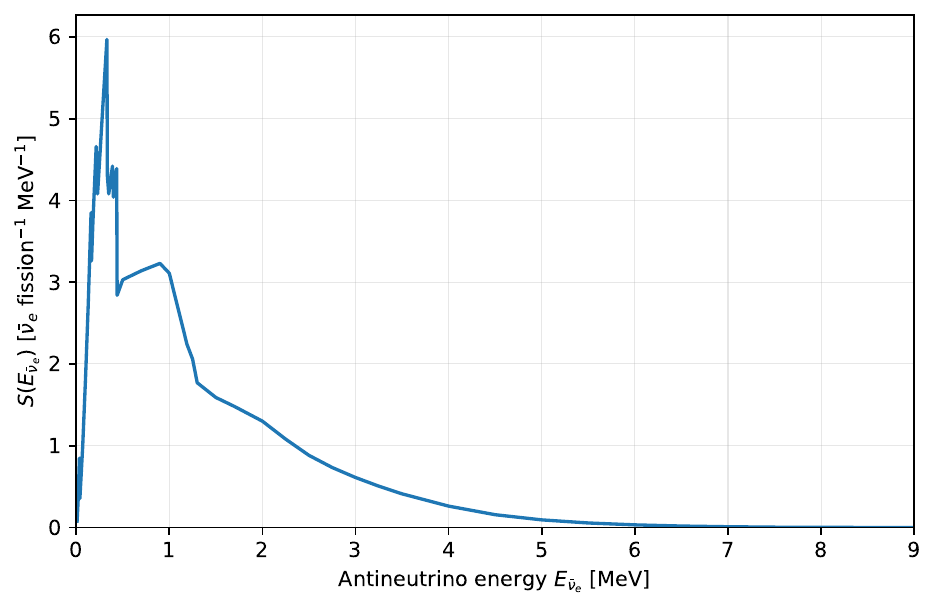}
        \caption{
            Reactor antineutrino yield per fission as a function of
            antineutrino energy.
        }
        \label{fig:reactor_antineutrino_spectrum}
    \end{subfigure}
    \hfill
    \begin{subfigure}[t]{0.49\textwidth}
        \centering
        \includegraphics[width=\linewidth]
        {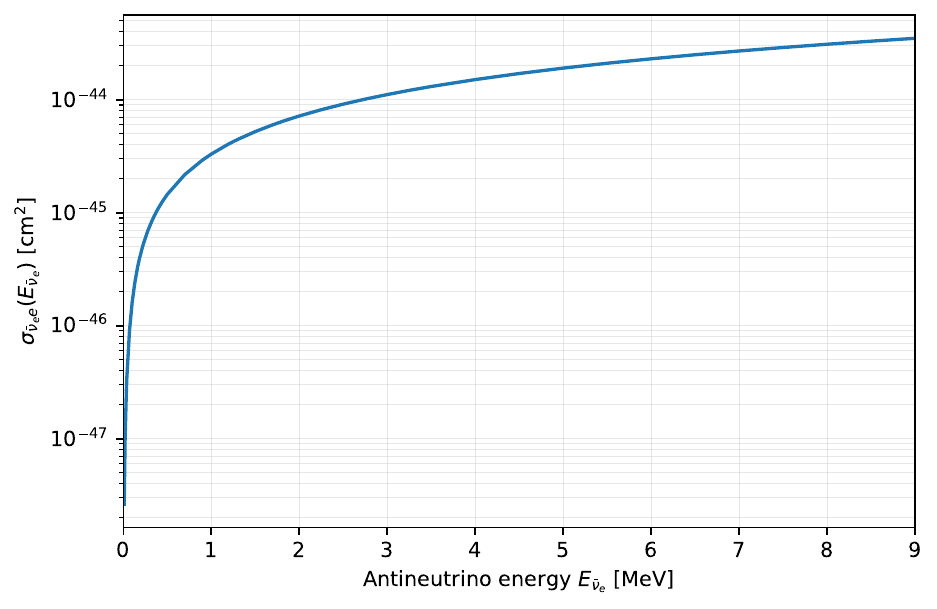}
        \caption{
            Energy-dependent total cross section for
            $\bar{\nu}_e$--electron elastic scattering, calculated
            following ref.~\cite{VogelEngel1989} using the analytic
            implementation in RAT-PAC~\cite{RatpacTwo2023}.
        }
        \label{fig:nuebar_e_cross_section}
    \end{subfigure}

    \caption{
        Reactor inputs used in the signal simulation and rate
        calculation.
    }
    \label{fig:reactor_inputs}
\end{figure}

The signal-rate calculation requires the reactor antineutrino spectrum
and the $\bar{\nu}_e$--electron elastic-scattering cross section.
The adopted differential reactor antineutrino yield per fission,
$S(E_{\bar{\nu}_e})$, shown in
figure~\ref{fig:reactor_antineutrino_spectrum}, is taken from
ref.~\cite{Kopeikin2012}.
The corresponding fission fractions are also adopted from the same reference. Numerical integration over the tabulated
energy range gives
\begin{equation}
    \frac{N_{\bar{\nu}_e}}{N_f}
    =
    \int_{\SI{0.01}{MeV}}^{\SI{9}{MeV}}
    S(E_{\bar{\nu}_e})\,\mathrm{d}E_{\bar{\nu}_e}
    =
    6.62.
    \label{eq:reactor_spectrum_yield}
\end{equation}

The corresponding elastic-scattering cross section, shown in
figure~\ref{fig:nuebar_e_cross_section}, is calculated following the
electroweak formalism of ref.~\cite{VogelEngel1989}, as implemented in
RAT-PAC~\cite{RatpacTwo2023}.  The differential cross section with
respect to the recoil-electron kinetic energy $T$ is
\begin{equation}
    \frac{\mathrm{d}\sigma_{\bar{\nu}_e e}}{\mathrm{d}T}
    =
    \frac{2G_{\mathrm F}^{2}m_e}{\pi}
    \left[
        g_L^{2}
        +
        g_R^{2}
        \left(
            1-\frac{T}{E_{\bar{\nu}_e}}
        \right)^{2}
        -
        g_L g_R
        \frac{m_e T}{E_{\bar{\nu}_e}^{2}}
    \right],
    \label{eq:nuebar_e_differential_cross_section}
\end{equation}
where the couplings for electron antineutrinos are defined, in the
convention used in the implementation, as
\begin{equation}
    g_L=\sin^{2}\theta_{\mathrm W},
    \qquad
    g_R=\frac{1}{2}+\sin^{2}\theta_{\mathrm W}.
    \label{eq:nuebar_e_couplings}
\end{equation}
The total cross section at a given antineutrino energy is then obtained
by integrating over the allowed recoil-electron kinetic-energy range:
\begin{equation}
    \sigma_{\bar{\nu}_e e}(E_{\bar{\nu}_e})
    =
    \int_{0}^{T_{\max}}
    \frac{\mathrm{d}\sigma_{\bar{\nu}_e e}}{\mathrm{d}T}
    \,\mathrm{d}T,
    \qquad
    T_{\max}
    =
    \frac{2E_{\bar{\nu}_e}^{2}}
         {m_e+2E_{\bar{\nu}_e}}.
    \label{eq:nuebar_e_total_cross_section}
\end{equation}

The implementation used in this study adopts
$\sin^{2}\theta_{\mathrm W}=0.232$,
$G_{\mathrm F}=1.17\times10^{-5}\,\mathrm{GeV^{-2}}$, and
$m_e=\SI{0.511}{MeV}$, with the resulting cross section converted
to $\mathrm{cm^2}$ using
$1\,\mathrm{GeV^{-2}}
=0.389\times10^{-27}\,\mathrm{cm^2}$.

Averaging \revgreen{the} total cross section over the \revgreen{adopted} reactor antineutrino
spectrum gives
\begin{equation}
    \left\langle \sigma_{\bar{\nu}_e e} \right\rangle
    =
    \frac{
        \displaystyle
        \int
        S(E_{\bar{\nu}_e})
        \sigma_{\bar{\nu}_e e}(E_{\bar{\nu}_e})
        \,\mathrm{d}E_{\bar{\nu}_e}
    }{
        \displaystyle
        \int
        S(E_{\bar{\nu}_e})
        \,\mathrm{d}E_{\bar{\nu}_e}
    }
    =
    4.82\times10^{-45}\,\mathrm{cm^2}.
    \label{eq:spectrum_averaged_es_cross_section}
\end{equation}

Combining the target-electron count, reactor fission rates,
antineutrino spectrum, and elastic-scattering cross section, and
assuming isotropic antineutrino emission from each reactor core, the
interaction rate from core $k$ is calculated as
\begin{equation}
    R_k
    =
    \frac{
        N_e \dot{N}_{f,k}
    }{
        4\pi L_k^2
    }
    \int
    S(E_{\bar{\nu}_e})
    \sigma_{\bar{\nu}_e e}(E_{\bar{\nu}_e})
    \,\mathrm{d}E_{\bar{\nu}_e},
    \label{eq:reactor_es_rate}
\end{equation}
where $L_k$ is the baseline between core $k$ and the detector.  The
resulting interaction rates for the three baseline groups are
summarized in table~\ref{tab:reactor_configuration}.

\begin{table}[t]
    \centering
    \caption{
        Expected reactor $\bar{\nu}_e$--electron elastic-scattering
        interaction rates in the \SI{170}{t} water target. The rates
        are calculated before detector response and event selection.
    }
    \label{tab:reactor_configuration}
    \begin{tabular}{lcccc}
        \toprule
        Reactor cores
            & Baseline
            & Number of cores
            & Rate per core
            & Subtotal \\
        &
            &
            & (\si{day^{-1}})
            & (\si{day^{-1}}) \\
        \midrule
        3 and 4
            & \SI{322}{m}
            & 2
            & $1.05\times10^{3}$
            & $2.09\times10^{3}$ \\
        2 and 5
            & \SI{488}{m}
            & 2
            & $4.56\times10^{2}$
            & $9.12\times10^{2}$ \\
        1 and 6
            & \SI{713}{m}
            & 2
            & $2.14\times10^{2}$
            & $4.27\times10^{2}$ \\
        \midrule
        Total
            & ---
            & 6
            & ---
            & $3.43\times10^{3}$ \\
        \bottomrule
    \end{tabular}
\end{table}

Summing the contributions from all six reactor cores gives a total
unoscillated interaction rate of approximately
$3.43\times10^{3}$ events per day.
If neutrino oscillations are taken into account, including both
the $\bar{\nu}_e$ disappearance and the elastic-scattering contribution
from oscillated $\bar{\nu}_{\mu}$ and $\bar{\nu}_{\tau}$, the resulting
change in the total interaction rate is expected to be of order 1\%.
This value represents the total interaction rate in the water target
before detector response, reconstruction requirements, and analysis
selections are applied.

\subsection{Expected background sources}
\label{subsec:expected_backgrounds}

Any process producing a reconstructed electron-like event in the water
can enter the selected sample.  The background model includes
radioactivity in the PMT glass, beta-decaying isotopes produced by
cosmic-ray muons, and beta decays of dissolved radon daughters.
A dedicated reactor-correlated IBD component is also considered as a
potential prompt background when the accompanying neutron is not identified.  This
subsection defines the nominal physical rates used to normalize the
corresponding Monte Carlo samples; the event-generation procedures are
described in section~\ref{subsec:background_generation}.  The adopted
rates are summarized in table~\ref{tab:background_rates}. These rates represent projections based on the assumptions of this
study; in an experimental implementation, their values are to be
determined from direct measurements of the selected detector materials
and the local operating environment.

\subsubsection{PMT radioactivity}

Radioactive contaminants in PMT glass emit gamma rays that can enter
the water and produce electron-like events through Compton scattering
and secondary electromagnetic interactions.  The components included
in the PMT-radioactivity model are $^{214}\mathrm{Bi}$ in the
$^{238}\mathrm{U}$ chain, $^{208}\mathrm{Tl}$ in the
$^{232}\mathrm{Th}$ chain, and $^{40}\mathrm{K}$.  Relevant gamma
lines include the \SI{609.3}{keV}, \SI{1120.3}{keV},
\SI{1764.5}{keV}, and \SI{2204}{keV} transitions from
$^{214}\mathrm{Bi}$, the \SI{2614.5}{keV} transition from
$^{208}\mathrm{Tl}$, and the \SI{1460.8}{keV} transition from
$^{40}\mathrm{K}$~\cite{BeRadionuclides2008}.

The PMT-radioactivity model uses the area-normalized activities
measured for the low-background Hamamatsu R11410-21 PMT as reference
inputs~\cite{AprilePMT2015}; 
differences in glass composition and material mass between the
R11410-21 and the reference 4-inch PMT are not modeled. The decay rate assigned to isotope $i$
in one PMT is calculated as
\begin{equation}
    r_i^{\mathrm{PMT}}
    =
    a_i A_{\mathrm{PMT}} f_i
    \left(
        \frac{86.4\ \mathrm{decays\,day^{-1}}}
             {1\ \mathrm{mBq}}
    \right),
    \label{eq:pmt_radioactivity_scaling}
\end{equation}
where $a_i$ is the activity per unit area,
$A_{\mathrm{PMT}}=\pi(D_{\mathrm{PMT}}/2)^2
=\SI{81.1}{cm^2}$ for $D_{\mathrm{PMT}}=\SI{101.6}{mm}$, and $f_i$
is the relevant decay-chain branching fraction.  Assuming secular
equilibrium, $f_{^{214}\mathrm{Bi}}=f_{^{40}\mathrm{K}}=1$ and
$f_{^{208}\mathrm{Tl}}=0.3594$.  Using
$a_{^{226}\mathrm{Ra}}=(0.016\pm0.003)\,\si{mBq.cm^{-2}}$,
$a_{^{228}\mathrm{Th}}=(0.012\pm0.003)\,\si{mBq.cm^{-2}}$, and
$a_{^{40}\mathrm{K}}
=(0.37\pm0.06)\,\si{mBq.cm^{-2}}$ gives
\begin{equation}
    \begin{aligned}
        r_{^{214}\mathrm{Bi}}^{\mathrm{PMT}}
            &=(113\pm21)
              \ \mathrm{day^{-1}\,PMT^{-1}},\\
        r_{^{208}\mathrm{Tl}}^{\mathrm{PMT}}
            &=(31\pm8)
              \ \mathrm{day^{-1}\,PMT^{-1}},\\
        r_{^{40}\mathrm{K}}^{\mathrm{PMT}}
            &=(2.60\pm0.42)\times10^{3}
              \ \mathrm{day^{-1}\,PMT^{-1}}.
    \end{aligned}
    \label{eq:pmt_decay_rates}
\end{equation}
The detector-wide rates are obtained from
$R_i^{\mathrm{det}}=N_{\mathrm{PMT}}r_i^{\mathrm{PMT}}$, with
$N_{\mathrm{PMT}}=9480$.

\subsubsection{Cosmogenic isotopes}

Cosmic-ray muons traversing the detector can produce unstable nuclei
through spallation reactions on oxygen and other nuclei in the water.
Subsequent beta decays generate single electron-like events over the
energy range relevant to the elastic-scattering analysis.  The
isotopes included in the background model are $^{12}\mathrm{B}$,
$^{12}\mathrm{N}$, $^{16}\mathrm{N}$, and $^{8}\mathrm{Li}$.  Their
production yields in water are taken from measurements in
Super-Kamiokande~\cite{ZhangSpallation2016} and rescaled using the muon
conditions at the RENO near-detector site.

A muon flux of
$(6.67\pm0.15)\,\si{m^{-2}.s^{-1}}$ and a mean muon energy of
$(33.1\pm2.3)\,\si{GeV}$ are adopted from
ref.~\cite{LeeRENO2022}.  The mean muon track length in the water
target is evaluated using the near-detector zenith-angle distribution
reported in the same reference, parameterized as a function of
$\mu=\cos\theta$. For a cylindrical
target with radius $R$ and height $H$, the projected area normal to
the muon direction is
\begin{equation}
    A_{\mathrm{proj}}(\mu)
    =
    \pi R^2\mu
    +
    2RH\sqrt{1-\mu^2},
    \label{eq:muon_projected_area}
\end{equation}
and the corresponding directional mean chord length is
\begin{equation}
    \overline{L}_{\mu}(\mu)
    =
    \frac{\pi R^2H}{A_{\mathrm{proj}}(\mu)}.
    \label{eq:muon_directional_track_length}
\end{equation}
Weighting this quantity by the digitized RENO near-detector angular
distribution, $w_{\mathrm{ND}}(\mu)$, gives
\begin{equation}
    \left\langle L_{\mu}\right\rangle
    =
    \frac{
        \displaystyle
        \int_0^1
        w_{\mathrm{ND}}(\mu)
        \overline{L}_{\mu}(\mu)
        \,\mathrm{d}\mu
    }{
        \displaystyle
        \int_0^1
        w_{\mathrm{ND}}(\mu)
        \,\mathrm{d}\mu
    }
    =
    \SI{4.015}{m},
    \label{eq:mean_muon_track_length}
\end{equation}
for $R=\SI{3}{m}$ and $H=\SI{6}{m}$.

The energy dependence of the isotope yield is parameterized as
\begin{equation}
    Y_i(E_\mu)
    =
    Y_i^{\mathrm{SK}}
    \left(
        \frac{E_\mu}{E_{\mu,\mathrm{SK}}}
    \right)^{\alpha_i},
    \label{eq:cosmogenic_scaling}
\end{equation}
with
$E_{\mu,\mathrm{SK}}=(259\pm9)\,\si{GeV}$~\cite{ShinokiNeutron2023}.
A common exponent $\alpha_i=0.75$ is adopted for all included isotopes,
consistent with the measured energy dependence of cosmogenic production
in liquid scintillator~\cite{AbeKamLAND2010,LeeRENO2022}. Under the flux
convention used here, the total rate of muons entering the cylindrical target
is approximated by
\begin{equation}
    R_{\mu}
    =
    \Phi_{\mu}\,\pi R^{2},
    \label{eq:detector_muon_rate}
\end{equation}
where $\Phi_{\mu}$ is the measured muon flux at the RENO near-detector site. This simplified treatment neglects the additional geometrical
acceptance for inclined muons entering through the lateral surface and is
adopted as part of the nominal background model for the expected-performance
study.
Since the production yield is quoted per muon and per target areal density, the
rate of isotope $i$ is then calculated as
\begin{equation}
    R_i
    =
    R_{\mu}\,\rho\,\left\langle L_{\mu}\right\rangle
    Y_i(E_{\mu}),
    \label{eq:cosmogenic_rate_calculation}
\end{equation}
with $\rho$ the water density and all lengths converted consistently so that
$\rho\langle L_{\mu}\rangle$ is expressed in
$\si{g.cm^{-2}}$. The quoted rate uncertainties propagate the measured muon
flux, mean muon energy, and reference isotope-yield uncertainties. The exponent
$\alpha_i=0.75$ and the calculated mean track length are treated as fixed in
the present projection. The resulting isotope-production rates are
\begin{equation}
    \begin{aligned}
        R_{^{12}\mathrm{B}}
            &=(1.60\pm0.13)\times10^{3}
              \ \si{day^{-1}},\\
        R_{^{12}\mathrm{N}}
            &=(2.00\pm0.22)\times10^{2}
              \ \si{day^{-1}},\\
        R_{^{16}\mathrm{N}}
            &=(3.30\pm0.41)\times10^{3}
              \ \si{day^{-1}},\\
        R_{^{8}\mathrm{Li}}
            &=(7.00\pm0.60)\times10^{2}
              \ \si{day^{-1}}.
    \end{aligned}
    \label{eq:cosmogenic_isotope_rates}
\end{equation}

\subsubsection{Water-borne radon}

Dissolved $^{222}\mathrm{Rn}$ and its daughters form an important
low-energy background in large water Cherenkov detectors. In the
decay sequence
\begin{equation}
    ^{222}\mathrm{Rn}\rightarrow{}^{218}\mathrm{Po}
    \rightarrow{}^{214}\mathrm{Pb}\rightarrow{}^{214}\mathrm{Bi}
    \rightarrow{}^{214}\mathrm{Po},
    \label{eq:radon_chain}
\end{equation}
the alpha particles from the radon and polonium decays are below the
Cherenkov threshold in water, whereas the beta decays of
$^{214}\mathrm{Pb}$ and $^{214}\mathrm{Bi}$ can produce detectable
electron-like events. The background model adopts a $^{222}\mathrm{Rn}$
concentration of \SI{0.14}{mBq.m^{-3}}, motivated by measurements of
highly purified water in Super-Kamiokande~\cite{NakanoRadon2020}.
The dissolved radon is assumed to be uniformly distributed
throughout the water target. For a \SI{170}{m^3} target, this
concentration corresponds to approximately $2.1\times10^3$ decays per
day. Under the assumption of secular equilibrium, the same decay rate
is assigned separately to $^{214}\mathrm{Pb}$ and $^{214}\mathrm{Bi}$
before reconstruction and event selection.

\begingroup
\subsubsection{Inverse beta decay}
\label{subsubsec:ibd_background}

Inverse beta decay,
$\bar{\nu}_{e}+p\rightarrow e^{+}+n$, has a substantially larger interaction
rate than antineutrino--electron elastic scattering at reactor energies. In a
water Cherenkov detector, an IBD event can enter the ES candidate sample if the
prompt positron satisfies the reconstruction and prompt-hit requirements while
the neutron is not identified. Although the IBD positron direction is only
weakly correlated with the incident antineutrino direction, the interaction
rate itself is reactor correlated.

The \SI{170}{t} water target contains
$N_p=(2/10)N_e=1.14\times10^{31}$ free proton targets. Using the
spectrum-integrated IBD yield per fission measured by RENO, 
$\sigma_f^{\mathrm{IBD}}=5.852\times10^{-43}\,\mathrm{cm^2\,fission^{-1}}$~\cite{RENOFlux2021},
together with the core fission rates and baselines defined in
section~\ref{subsec:reactor_configuration}, gives a nominal all-core unoscillated IBD
interaction rate of approximately
$1.26\times10^{4}\,\mathrm{day^{-1}}$ before detector response and event
selection. This rate is used to normalize the untagged IBD Monte Carlo sample.

The neutron can in principle provide a delayed tag, particularly in
Gd-loaded water~\cite{Hellfeld2017}. Because neutron thermalization, capture,
and the resulting delayed-coincidence efficiency are not modeled in the
present reference detector simulation, no specific Gd concentration or
neutron-tagging performance is assumed. Instead, the impact of residual IBD
contamination is evaluated parametrically in
section~\ref{subsec:ibd_tagging_impact} for several assumed tagging
efficiencies.
\endgroup

\begin{table}[t]
    \centering
    \caption{
    Nominal physical background rates used to normalize the Monte
    Carlo samples.
}
    \label{tab:background_rates}
    \small
    \begin{tabular}{lc}
        \toprule
        Component & Nominal physical rate \\
        \midrule
        PMT $^{214}\mathrm{Bi}$
            & $(113\pm21)\ \mathrm{day^{-1}\,PMT^{-1}}$ \\
        PMT $^{208}\mathrm{Tl}$
            & $(31\pm8)\ \mathrm{day^{-1}\,PMT^{-1}}$ \\
        PMT $^{40}\mathrm{K}$
            & $(2.60\pm0.42)\times10^{3}
              \ \mathrm{day^{-1}\,PMT^{-1}}$ \\
        \addlinespace
        $^{12}\mathrm{B}$
            & $(1.60\pm0.13)\times10^{3}
              \ \mathrm{day^{-1}}$ \\
        $^{12}\mathrm{N}$
            & $(2.00\pm0.22)\times10^{2}
              \ \mathrm{day^{-1}}$ \\
        $^{16}\mathrm{N}$
            & $(3.30\pm0.41)\times10^{3}
              \ \mathrm{day^{-1}}$ \\
        $^{8}\mathrm{Li}$
            & $(7.00\pm0.60)\times10^{2}
              \ \mathrm{day^{-1}}$ \\
        \addlinespace
        Water $^{214}\mathrm{Pb}$
            & $2.1\times10^{3}\ \mathrm{day^{-1}}$ \\
        Water $^{214}\mathrm{Bi}$
            & $2.1\times10^{3}\ \mathrm{day^{-1}}$ \\
        \addlinespace
        {Reactor IBD}
            & {$1.26\times10^{4}\ \mathrm{day^{-1}}$} \\
        \bottomrule
    \end{tabular}
\end{table}
\section{Monte Carlo simulation}
\label{sec:simulation}

The signal and background samples used in this study were generated with
RAT-PAC~\cite{RatpacTwo2023,RatpacDocs2026}.  The detector geometry and
material properties described in
section~\ref{sec:detector_conditions} were implemented in the simulation,
and separate source configurations were used for the signal and
background components.  Particle interactions, optical-photon production
and transport, PMT response, event triggering, and waveform digitization
were simulated within the same framework.  The following subsections
describe the detector-response configuration and the generation of the
reactor antineutrino signal and background samples.  The simulated sample
sizes were independent of the expected physical event rates, which were
applied after reconstruction.

\subsection{Detector and optical simulation}
\label{subsec:detector_simulation}

All signal and background samples used the detector geometry described in
section~\ref{subsec:detector_geometry}.  The target medium was pure water
at a nominal temperature of \SI{295}{\kelvin} and a density of
\SI{1000}{\kilogram\per\cubic\metre}.  Electromagnetic interactions of electrons,
positrons, photons, and radioactive-decay products were simulated
together with the resulting optical-photon production and transport.

Cherenkov photons were generated using the Geant4 Cherenkov process as
implemented in RAT-PAC~\cite{RatpacTwo2023,RatpacDocs2026}.  Their
propagation through the water was determined by the wavelength-dependent
refractive index, absorption length, and Rayleigh-scattering length.  The
refractive index was calculated using the IAPWS formulation under the
nominal water conditions described above~\cite{IAPWS1997}, while the
absorption and Rayleigh-scattering lengths were obtained from
refs.~\cite{FewellTrojan2019,ZhangHu2021}, respectively.  The optical
properties used in the simulation are summarized in
appendix~\ref{app:water_optical_properties}.

When an optical photon reached a PMT, its probability of producing a
photoelectron was determined from the wavelength-dependent photocathode
efficiency, including corrections for the incidence angle and
polarization.  The charge and transit time of each detected photoelectron
were then sampled from the corresponding single-photoelectron response
distributions.

Each PMT hit contributed a \SI{10}{ns}-wide trigger pulse, and an event
trigger was issued when the global trigger sum reached a threshold of 8.
For each trigger, a \SI{600}{ns} event window was recorded, including a
\SI{200}{ns} lookback period.  A \SI{400}{ns} lockout was applied
independently to each PMT.  Dark noise and PMT afterpulsing were not
included in the simulation.

\subsection{Reactor antineutrino--electron scattering signal generation}
\label{subsec:signal_generation}

The reactor antineutrino signal was generated through elastic scattering
of antineutrinos on electrons in water. Six source components were
simulated, corresponding to the six Hanbit reactor cores. Interaction
vertices were sampled uniformly throughout the water target, and the
incident antineutrino direction for each component was defined from the
corresponding reactor core toward the detector.

For each event, the incident antineutrino energy was sampled from the
reactor antineutrino spectrum based on ref.~\cite{Kopeikin2012}, as
introduced in section~\ref{subsec:reactor_configuration}. The recoil
electron kinematics were sampled according to the differential
$\bar{\nu}_e$--electron scattering cross section described in the same
section. The generated electron therefore retained the directional
correlation with the corresponding reactor core, which provides the
principal observable used in the analysis.

The six reactor-core components were combined with relative probabilities
proportional to the inverse square of their baselines,
\begin{equation}
    w_k
    =
    \frac{L_k^{-2}}{\sum_{j=1}^{6}L_j^{-2}},
\end{equation}
where $L_k$ is the distance from core $k$ to the detector. A total of
$1.0\times10^{5}$ antineutrino--electron elastic-scattering events was
generated in the combined six-core sample. The physical event-rate
normalization was applied after reconstruction using the rates given in
table~\ref{tab:reactor_configuration}.

\subsection{Background event generation}
\label{subsec:background_generation}

Each radioactive background component was generated in an independent
Monte Carlo sample. PMT radioactivity was distributed within the PMT
glass, whereas cosmogenic isotopes and radon-chain daughters were
distributed throughout the water target. The radioactive-decay
branches and emitted-particle spectra used in the simulation were taken
from the decay data implemented in RAT-PAC~\cite{RatpacTwo2023,RatpacDocs2026}.
After reconstruction, the samples were normalized to the expected decay
rates listed in table~\ref{tab:background_rates}.

The PMT-radioactivity background included
$^{214}\mathrm{Bi}$, $^{208}\mathrm{Tl}$, and $^{40}\mathrm{K}$ decays
generated uniformly within the PMT glass volumes. Although these
components are categorized as PMT-gamma backgrounds, the events passing
the analysis selection are predominantly initiated by photons escaping
from the PMT glass into the water and undergoing subsequent
electromagnetic interactions. A sample of $1.0\times10^{6}$ decays was
generated independently for each isotope.

The cosmogenic-isotope background included
$^{12}\mathrm{B}$, $^{12}\mathrm{N}$, $^{16}\mathrm{N}$, and
$^{8}\mathrm{Li}$. The decay vertices of all four isotopes were
distributed uniformly throughout the water target. Only the decays of
the produced isotopes were simulated; the parent cosmic-ray muons and
the associated spallation processes were not included. Consequently,
the samples do not contain event-by-event correlations with preceding
muon tracks or the timing information required to associate the decays
with a muon-veto window. The $^{12}\mathrm{N}$ sample represents the
dominant ground-state $\beta^+$ branch, whose branching fraction is
0.943944, and the remaining 5.61\% of its decays were not included in
the background model. A total of $1.0\times10^{5}$ decays was generated
independently for each isotope.

The water-borne radon background was represented by independent
$^{214}\mathrm{Pb}$ and $^{214}\mathrm{Bi}$ decay samples. Their decay
vertices were distributed uniformly throughout the water target. The two
isotopes were simulated independently rather than as a correlated
$^{222}\mathrm{Rn}$ decay sequence, and no delayed-coincidence relation
between the $^{214}\mathrm{Pb}$ and $^{214}\mathrm{Bi}$ decays was
included. A sample of $1.0\times10^{5}$ decays was generated
independently for each isotope.

A dedicated sample of $1.0\times10^{5}$ IBD events was also generated with the RAT-PAC IBD vertex generator. The same reactor spectrum, core directions, and inverse-square baseline weighting used for the ES sample were adopted. The simulated IBD events were processed through the same reconstruction and event-selection procedure as the other background samples, and the resulting prompt-event sample was used to evaluate IBD contamination in the ES candidate sample. Neutron tagging was not simulated explicitly; instead, its efficiency was treated as an assumed parameter in the sensitivity study. The physical normalization was applied after reconstruction using the nominal IBD rate given in table~\ref{tab:background_rates}.

\section{Event reconstruction and selection}
\label{sec:reconstruction}

The event reconstruction was developed to retain the directional information
carried by Cherenkov photons emitted by the recoil electron.  The procedure
consists of three stages: prompt-hit selection, time-of-flight-based vertex
reconstruction, and spherical Hough-transform reconstruction of the electron
direction.  The reconstruction uses the PMT hit positions, hit times, and
photoelectron counts.

\subsection{Hit preprocessing and prompt-hit selection}
\label{subsec:preselection}

Multiple photoelectrons detected by the same PMT were combined into a single
PMT entry.  The PMT position was used as the channel identifier, the earliest
photoelectron time was assigned as the representative hit time $t_i$, and the
number of photoelectrons was retained as the charge-like quantity $q_i$.  A
nominal detector timing resolution of \SI{1}{ns} was assumed for the recorded
hit times.  Events containing fewer than 8 hit PMTs were rejected before
reconstruction.

A prompt-hit set was formed relative to the earliest PMT time in each event,
\begin{equation}
    \Delta t_i = t_i - t_{\min},
    \label{eq:prompt_time}
\end{equation}
where $t_{\min}$ is the earliest representative PMT hit time.  Events were
required to contain at least 8 distinct PMTs satisfying
\begin{equation}
    \Delta t_i < \SI{10}{ns}.
    \label{eq:prompt_selection}
\end{equation}
The number of PMTs satisfying this condition is denoted
$N_{\mathrm{prompt}}$.  Only these prompt hits were used in the vertex fit.
This selection suppresses delayed photons produced by scattering or
reflections while retaining the compact direct-light timing peak.

\subsection{Vertex reconstruction}
\label{subsec:vertex_reconstruction}

The vertex reconstruction was adapted from the timing-goodness method
developed for Super-Kamiokande~\cite{Shiozawa1999,Takenaka2020}.  In this
approach, the interaction vertex is identified as the position for which the
PMT hit-time distribution becomes most sharply concentrated after correcting
for the photon time of flight.  The present implementation follows the same
timing-goodness principle, while using a detector-specific timing model and
numerical search procedure.

For a candidate vertex $\mathbf{x}$, the time-of-flight-corrected hit time of
PMT $i$ was defined as
\begin{equation}
    t_i'(\mathbf{x})
    =
    (t_i-t_{\min})
    -
    \frac{|\mathbf{r}_i-\mathbf{x}|}{c_{\mathrm{w}}},
    \label{eq:tof_subtracted_time}
\end{equation}
where $\mathbf{r}_i$ is the PMT position and
$c_{\mathrm{w}}=c/n_{\mathrm{w}}$ is the speed of light in water.  The
refractive index was evaluated at a reference photon energy of \SI{3}{eV},
giving $n_{\mathrm{w}}=1.34$ and
$c_{\mathrm{w}}=\SI{22.3}{cm/ns}$.

The effective timing uncertainty assigned to each PMT was parametrized as
\begin{equation}
    \sigma_i
    =
    \max\left(
        \SI{1.5}{ns},
        \frac{\SI{5.0}{ns}}{\sqrt{q_i}}
    \right),
    \qquad
    w_i = \frac{1}{\sigma_i^2},
    \label{eq:vertex_timing_weight}
\end{equation}
where $q_i$ is the number of photoelectrons detected by PMT $i$.  In this
effective timing model, PMTs with larger photoelectron counts are assigned
smaller timing uncertainties and therefore contribute more strongly to the
vertex fit.  The lower bound of \SI{1.5}{ns} prevents the effective uncertainty
from becoming unrealistically small at large $q_i$, while the inverse-variance
weight $w_i$ gives greater importance to PMTs with more precise timing
information.

For each candidate vertex, the event-time offset was estimated from the
weighted mean of the time-of-flight-corrected hit times,
\begin{equation}
    \hat{t}_0(\mathbf{x})
    =
    \frac{\sum_i w_i t_i'(\mathbf{x})}
         {\sum_i w_i}.
    \label{eq:vertex_event_time}
\end{equation}
The quantity $\hat{t}_0(\mathbf{x})$ represents the common corrected-time
estimate associated with the candidate vertex.  For a candidate position near
the interaction vertex, the photon time-of-flight correction causes the
corrected hit times to cluster around this common value.

Following the Super-Kamiokande timing-goodness construction, the vertex
goodness was evaluated as
\begin{equation}
    G(\mathbf{x})
    =
    \frac{1}{\sum_i w_i}
    \sum_i w_i
    \exp\left[
        -\frac{
            \left(
                t_i'(\mathbf{x})-\hat{t}_0(\mathbf{x})
            \right)^2
        }{
            2\left(1.5\,\bar{\sigma}\right)^2
        }
    \right],
    \label{eq:vertex_score}
\end{equation}
where $\bar{\sigma}$ is the mean timing uncertainty of the selected PMTs.  The
exponential term gives a large contribution to PMTs whose corrected hit times
are consistent with the common timing peak and suppresses hits farther from
it.  The normalization by $\sum_i w_i$ expresses $G$ as a weighted average
rather than allowing it to increase solely with the number of selected hits.
A candidate position near the interaction vertex therefore produces a
narrower corrected-time distribution and a larger value of $G$.

The goodness was maximized using a multistage stochastic search restricted to
the cylindrical water volume.  The initial search position was defined as the
photoelectron-weighted centroid of the selected PMTs.  Candidate vertices were
then sampled around the best position identified at each stage, with the
search region progressively reduced to refine the vertex estimate.  The
position maximizing $G(\mathbf{x})$ was defined as the reconstructed vertex
$\hat{\mathbf{x}}$, and the corresponding
$\hat{t}_0(\hat{\mathbf{x}})$ was retained as the reconstructed event-time
offset.

A reconstructed fiducial-volume requirement was subsequently applied.  For
the cylindrical water volume, the distance from the reconstructed vertex to
the nearest detector boundary was defined as
\begin{equation}
    d_{\mathrm{wall}}(\hat{\mathbf{x}})
    =
    \min\left[
        R-\sqrt{\hat{x}^{2}+\hat{y}^{2}},
        Z-|\hat{z}|
    \right],
    \label{eq:reco_wall_distance}
\end{equation}
where $R,Z=\SI{300}{cm}$.  Events satisfying
\begin{equation}
    d_{\mathrm{wall}}(\hat{\mathbf{x}}) < \SI{100}{cm}
    \label{eq:reco_fiducial_cut}
\end{equation}
were excluded.  Defining the fiducial requirement in terms of the reconstructed
vertex allows the same selection to be applied to both simulated and
experimental data without relying on Monte Carlo truth information.

\subsection{Direction reconstruction}
\label{subsec:direction_reconstruction}

Using the reconstructed vertex and event-time offset, the recoil-electron
direction was reconstructed from the Cherenkov-light pattern using a spherical
Hough-transform method inspired by the ring-finding procedure developed for
Super-Kamiokande~\cite{Shiozawa1999}.  The algorithm searches for the
directional axis whose associated Cherenkov cone is most consistent with the
observed PMT hit pattern.

Before performing the directional fit, a set of hits broadly compatible with
direct Cherenkov light was selected.  For each PMT, the residual time was
defined as
\begin{equation}
    \delta t_i
    =
    t_i'(\hat{\mathbf{x}})
    -
    \hat{t}_0(\hat{\mathbf{x}}),
    \label{eq:direction_time_residual}
\end{equation}
where $t_i'(\hat{\mathbf{x}})$ is the time-of-flight-corrected hit time
evaluated at the reconstructed vertex.  PMTs satisfying
\begin{equation}
    |\delta t_i| < \SI{8}{ns}
    \label{eq:direction_tof_selection}
\end{equation}
were retained for the directional reconstruction.  This relatively loose
window removes clearly delayed PMT hits while retaining hits associated with
the prompt Cherenkov-light pattern.  Timing information was used only to
select the direction-fit hit set and was not included directly in the Hough
score.  When fewer than 8 PMTs satisfied the residual-time requirement, the
prompt-hit set used for the vertex reconstruction was used instead.

For each selected PMT, a unit vector pointing from the reconstructed vertex
toward the PMT was defined as
\begin{equation}
    \hat{\mathbf{r}}_i
    =
    \frac{\mathbf{r}_i-\hat{\mathbf{x}}}
         {|\mathbf{r}_i-\hat{\mathbf{x}}|}.
    \label{eq:pmt_direction_vector}
\end{equation}
For a candidate electron direction $\mathbf{u}$, a direct Cherenkov photon is
expected to satisfy
\begin{equation}
    \hat{\mathbf{r}}_i\cdot\mathbf{u}
    \simeq
    \cos\theta_{\mathrm{C}},
    \label{eq:cherenkov_cone_condition}
\end{equation}
where $\theta_{\mathrm{C}}$ is the Cherenkov angle in water.  Assuming a
relativistic electron with $\beta\simeq1$, this angle was calculated as
\begin{equation}
    \theta_{\mathrm{C}}
    =
    \cos^{-1}\left(\frac{1}{n_{\mathrm{w}}}\right),
    \label{eq:cherenkov_angle}
\end{equation}
using the same refractive index $n_{\mathrm{w}}$ as in the vertex
reconstruction.

The consistency of a candidate direction with the observed Cherenkov pattern
was quantified by the Hough score
\begin{equation}
    H(\mathbf{u})
    =
    \frac{1}{\sum_i a_i}
    \sum_i a_i
    \exp\left[
        -\frac{1}{2}
        \left(
            \frac{
                \hat{\mathbf{r}}_i\cdot\mathbf{u}
                -\cos\theta_{\mathrm{C}}
            }{
                \sin\theta_{\mathrm{C}}\,\sigma_{\theta}
            }
        \right)^2
    \right],
    \label{eq:direction_hough_score}
\end{equation}
with $a_i = \sqrt{q_i},$ where the value $\sigma_{\theta}=\SI{6}{\degree}$ is used as a fixed Hough-kernel width and should not be interpreted as the detector angular resolution.
The Gaussian term gives a large contribution to PMTs located near the expected
Cherenkov cone and suppresses PMTs farther from it.  The width
$\sigma_{\theta}$ accounts phenomenologically for the angular spread arising
from vertex uncertainty, electron scattering, and detector granularity.  The
weight $a_i=\sqrt{q_i}$ gives moderately greater importance to PMTs with larger
photoelectron counts while preventing individual high-charge PMTs from
dominating the fit.  Normalization by the total weight prevents the scale of
the score from increasing trivially with the number or charge of the selected
hits.

The reconstructed recoil-electron direction was defined as the direction
maximizing the Hough score,
\begin{equation}
    \hat{\mathbf{u}}
    =
    \operatorname*{arg\,max}_{|\mathbf{u}|=1}
    H(\mathbf{u}).
    \label{eq:reconstructed_direction}
\end{equation}
Candidate directions were initially distributed approximately uniformly over
the unit sphere. The direction with the largest Hough score was subsequently
refined through local searches with progressively reduced angular ranges.

\subsection{Reconstruction performance}
\label{subsec:reconstruction_performance}

The numerical thresholds introduced above were adopted as a baseline working
point for characterizing the overall reconstruction performance.  In
particular, the requirement of at least 8 prompt PMTs within \SI{10}{ns} and
the use of an \SI{8}{ns} residual-time window were not optimized as final
physics-analysis cuts.  In an experimental analysis, these requirements should
be adjusted according to the physics objective, detector background
conditions, reconstruction efficiency, and desired signal purity.

The baseline reconstruction performance was evaluated using the
$1.0\times10^{5}$ simulated antineutrino--electron scattering events described
in section ~\ref{subsec:signal_generation}.  Monte Carlo truth information was
used only to evaluate the reconstruction errors and was not used in the
reconstruction or event selection.  The selection counts for the baseline
configuration are summarized in
Table~\ref{tab:reconstruction_cutflow}.

Of the generated events, 75,871 contained at least one detected PMT hit.
After the distinct-PMT and prompt-hit requirements, 42,639 events entered the
vertex reconstruction.  A total of 14,361 events satisfied the complete
reconstruction and the reconstructed \SI{1}{m} fiducial requirement,
corresponding to an overall baseline selection efficiency of 14.36\% relative
to the generated sample.

\begin{table}[t]
    \centering
    \caption{Selection counts for the simulated antineutrino--electron
    scattering sample under the baseline reconstruction configuration.  The
    final row includes successful vertex and direction reconstruction together
    with the reconstructed \SI{1}{m} fiducial requirement.}
    \label{tab:reconstruction_cutflow}
    \begin{tabular}{lrr}
        \toprule
        Selection stage & Events & Fraction of generated sample \\
        \midrule
        Generated ES events
            & 100,000 & 100.00\% \\
        At least one detected PMT hit
            & 75,871 & 75.87\% \\
        Distinct-PMT and prompt-hit selection
            & 42,639 & 42.64\% \\
        Final reconstructed and fiducial sample
            & 14,361 & 14.36\% \\
        \bottomrule
    \end{tabular}
\end{table}

The vertex error was defined as
\begin{equation}
    \Delta r
    =
    |\hat{\mathbf{x}}-\mathbf{x}_{\mathrm{true}}|.
    \label{eq:vertex_error}
\end{equation}
For the final baseline-selected ES sample, the median vertex error was
\SI{15.33}{cm}, the 68\% containment was \SI{22.00}{cm}, and the root-mean-square
error was \SI{33.95}{cm}.

\begin{figure}[t]
    \centering
    \begin{subfigure}[t]{0.48\textwidth}
        \centering
        \includegraphics[width=\textwidth]
        {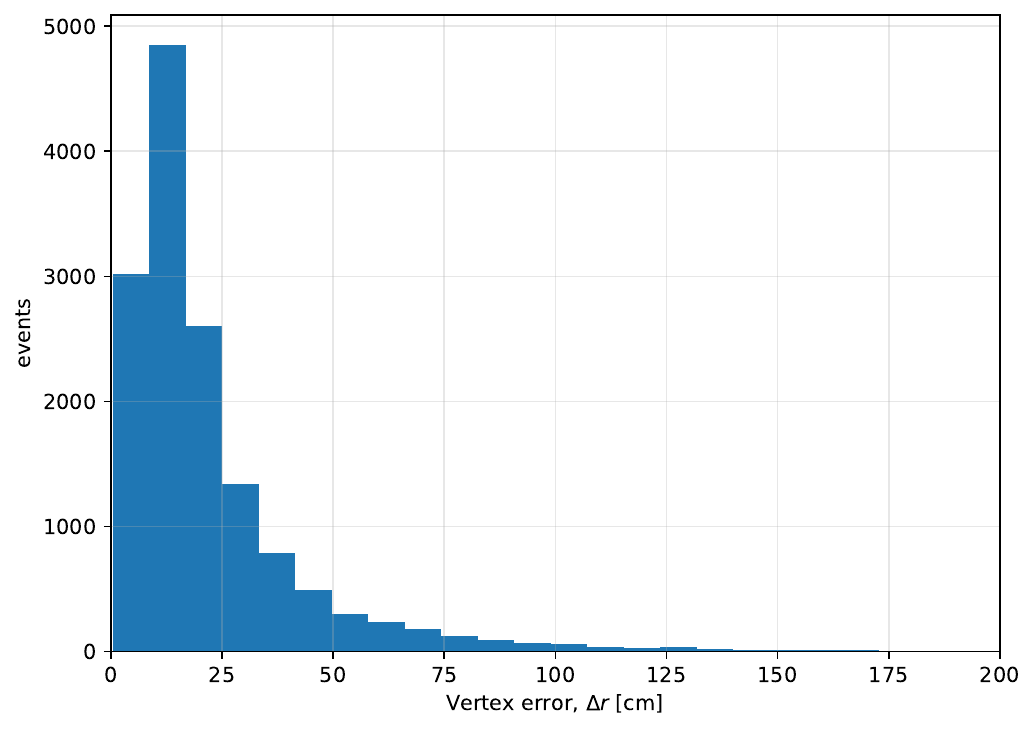}
        \caption{Distance between the reconstructed and true interaction
        vertices.}
        \label{fig:vertex_error_hough}
    \end{subfigure}
    \hfill
    \begin{subfigure}[t]{0.48\textwidth}
        \centering
        \includegraphics[width=\textwidth]
        {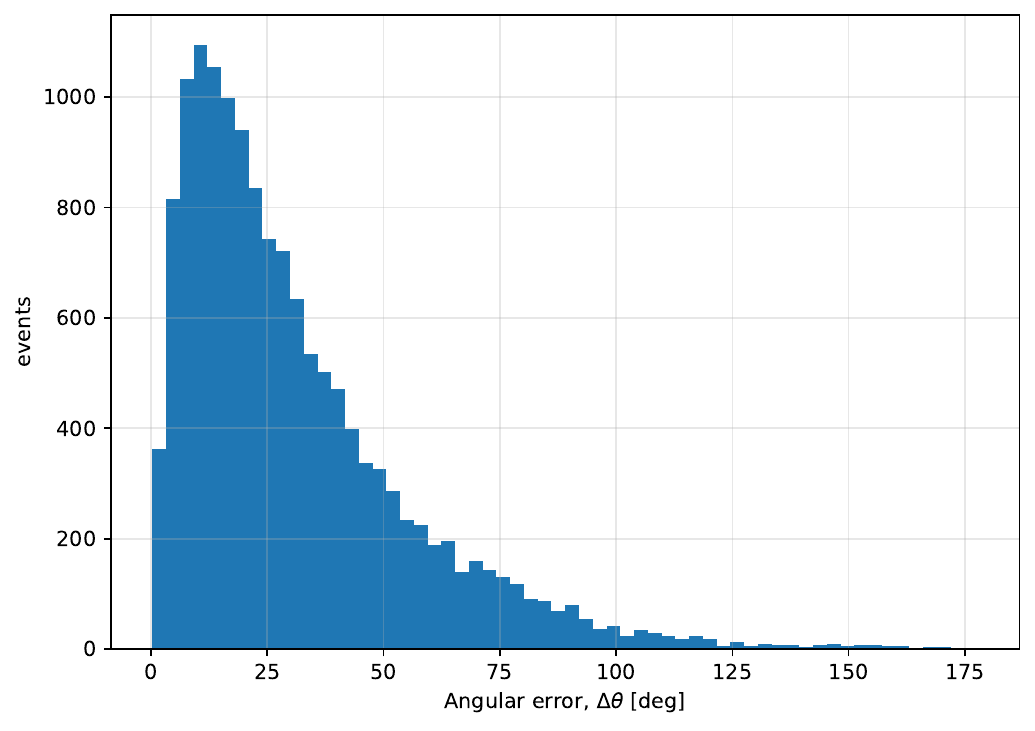}
        \caption{Angle between the Hough-reconstructed and true recoil-electron
        directions.}
        \label{fig:direction_error_hough}
    \end{subfigure}
    \caption{Vertex and direction reconstruction performance for the baseline
    selected antineutrino--electron scattering sample.}
    \label{fig:reconstruction_performance}
\end{figure}

The angular error was defined as
\begin{equation}
    \Delta\theta
    =
    \arccos\left(
        \hat{\mathbf{u}}\cdot\mathbf{u}_{\mathrm{true}}
    \right).
    \label{eq:direction_error}
\end{equation}
The Hough reconstruction gave a median angular error of
\SI{24.15}{\degree}, a 68\% containment of \SI{35.79}{\degree}, and an RMS
error of \SI{39.95}{\degree}.  The corresponding vertex- and angular-error
distributions are shown in figure ~\ref{fig:reconstruction_performance}.  The
complete angular-error distribution was retained when constructing the
reconstructed signal template in section~\ref{sec:analysis_performance}.

\section{Directional elastic-scattering analysis and projected performance}
\label{sec:analysis_performance}

This section presents the projected performance of the directional
$\bar{\nu}_{e}$--electron elastic-scattering (ES) analysis for an exposure of
365.25 days. The reconstructed recoil-electron direction is used to separate
the forward-peaked ES signal from the background. Two minimum prompt-hit
multiplicity requirements, $N_{\mathrm{prompt}}\geq8$ and
$N_{\mathrm{prompt}}\geq20$, are considered. Results are first presented for
the full selected energy range and are then divided into the idealized
$T_{e}^{\mathrm{true}}\leq3~\mathrm{MeV}$ and
$T_{e}^{\mathrm{true}}\geq3~\mathrm{MeV}$ regions. Here,
$T_{e}^{\mathrm{true}}$ denotes the kinetic energy of the primary particle in the RAT-PAC simulation. For the ES signal this variable is the recoil-electron
kinetic energy, while for the simulated backgrounds it corresponds to the
primary generated particle in nearly all events.

\subsection{Analysis strategy and directional observable}
\label{subsec:analysis_strategy}

\begin{figure}[t]
    \centering
    \includegraphics[width=0.78\textwidth]
    {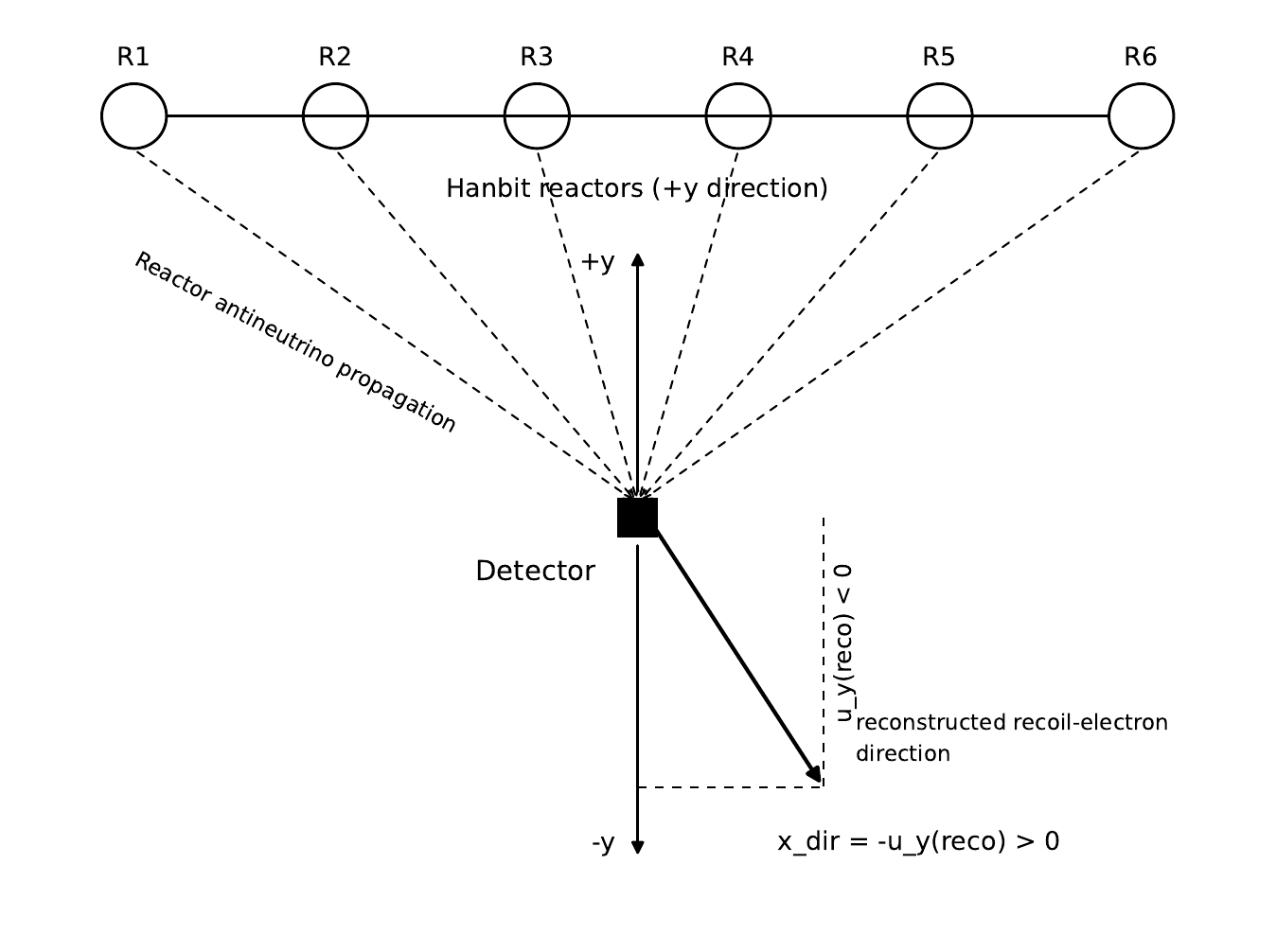}
    \caption{
        Schematic definition of the directional observable used in the
        $\bar{\nu}_{e}$--electron elastic-scattering analysis.
    }
    \label{fig:directional_observable}
\end{figure}

The reconstructed direction of the recoil electron is represented by the
unit vector
\begin{equation}
    \hat{\mathbf{u}}_{e}^{\mathrm{reco}}
    =
    \left(
        u_{x}^{\mathrm{reco}},
        u_{y}^{\mathrm{reco}},
        u_{z}^{\mathrm{reco}}
    \right).
\end{equation}
In the detector coordinate convention used in the simulation, the direction
from the detector toward the midpoint between reactor cores 3 and 4 corresponds to the positive
$y$ direction. Because the recoil electron tends to be emitted along the incident-antineutrino direction, signal events
tend to have $u_{y}^{\mathrm{reco}}\simeq-1$. The directional observable used
in the fit is consequently defined as
\begin{equation}
    x_{\mathrm{dir}}
    \equiv
    -u_{y}^{\mathrm{reco}}
    =
    -\hat{\mathbf{u}}_{e}^{\mathrm{reco}}
     \cdot\hat{\mathbf{y}}.
    \label{eq:directional_observable}
\end{equation}
With this convention, electrons reconstructed along the antineutrino
propagation direction, away from the reactors, have
$x_{\mathrm{dir}}\simeq1$. As illustrated in
figure~\ref{fig:directional_observable}, the forward-peaked ES signal therefore
produces an excess near $x_{\mathrm{dir}}=1$, while backgrounds with little
correlation with the antineutrino direction exhibit broader distributions.

The prompt-hit multiplicity $N_{\mathrm{prompt}}$ is the number of PMTs
registering a hit within the prompt time window used for event selection.
Increasing the requirement from $N_{\mathrm{prompt}}\geq8$ to
$N_{\mathrm{prompt}}\geq20$ removes events with fewer detected photons and
therefore defines a more restrictive selection. The two requirements are
treated as separate analysis scenarios; their final optimization for
experimental data is beyond the scope of this study.

The background composition also differs between the two scenarios. For
$N_{\mathrm{prompt}}\geq8$, the PMT-radioactivity components from
$^{40}\mathrm{K}$, $^{208}\mathrm{Tl}$, and $^{214}\mathrm{Bi}$ are not
included in the background model. Of the $10^{6}$ simulated $^{40}\mathrm{K}$ decays, only approximately 60
survive the reconstruction and selection. This limited surviving sample does
not provide sufficient statistical precision to reliably estimate the
selection efficiency and directional template, particularly given the very
large physical $^{40}\mathrm{K}$ activity. The small number of surviving events also leads to appreciable Monte Carlo fluctuations in the reconstructed directional template, preventing a robust quantitative treatment of this component. Generating substantially larger PMT-radioactivity samples was computationally impractical in the available simulation environment. The
$N_{\mathrm{prompt}}\geq8$ result is therefore presented as an idealized reference projection that neglects PMT radioactivity.

For $N_{\mathrm{prompt}}\geq20$, the PMT-radioactivity sources are included
in the background model together with the cosmogenic-isotope and
water-borne radon components. In the simulated $^{40}$K sample, no event
survives for thresholds at or above $N_{\mathrm{prompt}}=13$.
The adopted requirement of
$N_{\mathrm{prompt}}\geq20$ is therefore chosen conservatively above this
observed turnoff. For $10^{6}$ simulated $^{40}$K events with zero survivors,
the corresponding one-sided 95\% confidence upper limit on the selection
efficiency is approximately $3.0\times10^{-6}$.
The central $^{40}$K contribution is consequently set to zero in the nominal
$N_{\mathrm{prompt}}\geq20$ projection.
This treatment should not be interpreted as
establishing an exactly vanishing $^{40}$K contribution; rather, it reflects
the absence of surviving events in the available Monte Carlo sample.
The $^{208}$Tl and $^{214}$Bi components are retained according to their
simulated selection efficiencies.

\subsection{Likelihood construction and nuisance parameters}
\label{subsec:likelihood}

The directional observable $x_{\mathrm{dir}}$ was divided into 50 uniform
bins over the range $-1 \leq x_{\mathrm{dir}} \leq 1$. The signal and
individual background templates were constructed from the corresponding
selected Monte Carlo histograms, normalized to unit area, and subsequently
scaled to their expected physical event yields. The nominal
expected-sensitivity calculation used the full selected ES Monte Carlo sample
and source-specific Monte Carlo background shapes.

The signal normalization is extracted with a binned constrained
profile-likelihood fit. Let $s_i$ be the nominal signal expectation in bin
$i$, and let $b_{ij}$ be the nominal expectation from background source $j$.
The expected number of events is written as
\begin{equation}
    \mu_i(\alpha,\boldsymbol{\theta})
    =
    \alpha s_i
    \prod_{k\in\mathcal{S}}
    (1+r_k)^{\theta_k}
    +
    \sum_j
    b_{ij}(1+r_j)^{\theta_j},
    \label{eq:profile_model_expectation}
\end{equation}
where $\alpha$ is the ES signal-strength parameter, $r_k$ and $r_j$ are
relative one-standard-deviation normalization uncertainties, and the
corresponding nuisance parameters are constrained by standard normal
auxiliary measurements. The log-normal response $(1+r)^{\theta}$ preserves
positive source normalizations, equals the nominal scale at $\theta=0$, and
produces a scale factor $1+r$ at $\theta=+1$.
For the residual-IBD study, the surviving IBD template is included as
an additional background source in the same sum over $j$ and is not tied to the
ES signal-strength parameter $\alpha$. Its directional shape is kept fixed
while its nominal normalization is changed between the assumed tagging
scenarios.

The minimized negative log-likelihood is the Poisson likelihood relative to
the saturated model plus the Gaussian constraint terms,
\begin{equation}
    \mathrm{NLL}(\alpha,\boldsymbol{\theta})
    =
    \sum_i
    \left[
        \mu_i-n_i+n_i\ln\left(\frac{n_i}{\mu_i}\right)
    \right]
    +
    \frac{1}{2}\sum_{\ell}\theta_{\ell}^2,
    \label{eq:profile_nll}
\end{equation}
with the logarithmic term defined to be zero when $n_i=0$. The primary
interval is obtained by profiling all active nuisance parameters and finding
the points satisfying
\begin{equation}
    2\Delta\mathrm{NLL}=1,
    \label{eq:profile_interval_definition}
\end{equation}
which defines the nominal 68\% interval for one parameter of interest. A
conditional statistical interval is also calculated with the nuisance
parameters fixed at their profiled best-fit values. The difference in
quadrature between the total and conditional statistical intervals is used
only as a diagnostic effective systematic component; the profiled interval
is the primary result.

The expected-sensitivity calculation uses an Asimov data set,
\begin{equation}
    n_i^{\mathrm{A}}
    =
    \mu_i(\alpha=1,\boldsymbol{\theta}=\mathbf{0}).
    \label{eq:asimov_definition}
\end{equation}
Consequently, a matched model is expected to return
$\hat{\alpha}=1$ and the nominal values used to generate the templates.
Statistical and systematic uncertainties nevertheless broaden the profile
likelihood through the Poisson curvature and nuisance-parameter profiling.
The statistical performance and interval coverage of the likelihood
implementation are validated using pseudo-experiment ensembles in
section~\ref{subsec:statistical_validation}.

\subsection{Reactor-flux normalization extraction}
\label{subsec:reactor_flux_extraction}

The directional fit determines the normalization of the selected ES signal,
parameterized by the signal-strength parameter $\alpha$. The expected ES
yield is proportional to
\begin{equation}
    N_{\mathrm{ES}}
    \propto
    \Phi_{\bar{\nu}_e}
    \left\langle\sigma_{\bar{\nu}_e e}\right\rangle
    N_{e}\,t\,\epsilon_{\mathrm{ES}},
\end{equation}
where $\Phi_{\bar{\nu}_e}$ is the reactor antineutrino flux,
$\left\langle\sigma_{\bar{\nu}_e e}\right\rangle$ is the
spectrum-averaged ES cross section, $N_e$ is the number of target electrons,
$t$ is the exposure, and $\epsilon_{\mathrm{ES}}$ is the signal-selection
efficiency. The ES event rate alone therefore constrains the product of the
reactor flux and the ES cross section rather than their absolute
normalizations independently.

In this study, the fitted signal normalization is interpreted as the overall
normalization of the assumed reactor-antineutrino spectrum by constraining the
spectrum-averaged ES cross section. The spectral shape $S(E_{\bar{\nu}_e})$,
the integrated antineutrino yield per fission, the nominal core powers, and the
core-to-detector baselines are held fixed to the reactor model described in
section~\ref{subsec:reactor_configuration}. For convenience, the fitted
normalization is expressed as
\begin{equation}
    \Phi_{\mathrm{fit}}
    =
    \hat{\alpha}\,\Phi_{0},
    \label{eq:flux_projection_definition}
\end{equation}
where
\begin{equation}
    \Phi_{0}
    =
    1.449\times10^{11}
    \,\mathrm{cm^{-2}\,s^{-1}}
\end{equation}
is the nominal flux normalization predicted by the fixed reactor model. The
spectrum-averaged cross section is constrained around the nominal value
\begin{equation}
    \left\langle\sigma_{\bar{\nu}_e e}\right\rangle_{0}
    =
    4.821\times10^{-45}
    \,\mathrm{cm^{2}},
\end{equation}
with a conservative relative theoretical uncertainty of 0.5\%. This value
is chosen to cover the approximately 0.2--0.4\% theoretical uncertainty
estimated for precision calculations of elastic neutrino--electron
scattering~\cite{TomalakHill2020}.

All reactor-flux projections presented below follow this conditional
definition. Their statistical and total intervals are constructed using the
likelihood procedure described in section~\ref{subsec:likelihood}; the total
interval profiles the cross-section, signal-efficiency, and
background-normalization nuisance parameters. The quoted reactor-flux normalization is therefore model dependent, as it is
determined within the adopted reactor-spectrum model with fixed spectral shape,
fission yield, core powers, and baselines, and subject to the external
cross-section constraint.

Reactor-operation uncertainties are not included as nuisance parameters in
the present projection. The nominal reactor configuration described above is
therefore treated as a fixed reference model. A coherent variation of the
reactor thermal power would approximately produce a corresponding fractional
change in the predicted flux normalization, whereas variations in fuel
composition and fission fractions could also modify the reactor-antineutrino
spectral shape. Quantifying these effects would require core- and
time-dependent reactor-operation and fuel-composition information and is
beyond the scope of the present sensitivity study.

\subsection{Uncertainty summary}
\label{subsec:uncertainty_summary}

Table~\ref{tab:analysis_uncertainties} summarizes the uncertainties and
fixed assumptions used in the reactor-flux-normalization projections.
Normalization uncertainties are incorporated directly into the likelihood as
constrained nuisance parameters and are profiled when determining the total
68\% interval. Finite Monte Carlo statistics are propagated through the
binomial uncertainty of each selected fraction. For each background source,
this contribution is combined in quadrature with the assigned physical-rate
uncertainty before the corresponding log-normal nuisance parameter is
constructed.

The present uncertainty model is intended to quantify the statistical and
normalization uncertainties included in the sensitivity projection rather
than to provide a complete experimental systematic-uncertainty budget.
Several reactor-, detector-, and reconstruction-related quantities are kept
fixed in the present analysis, as summarized explicitly in
table~\ref{tab:analysis_uncertainties}.

The finite-Monte-Carlo efficiency term is a technical uncertainty associated
with the available simulated sample size rather than an intrinsic detector
limitation. For the all-energy ES selections, its relative size is 0.77\% for
$N_{\mathrm{prompt}}\geq8$ and 1.11\% for
$N_{\mathrm{prompt}}\geq20$. These contributions are retained because larger
samples are not available in the present computational environment; they would
decrease with additional simulation.

\begin{table}[t]
    \centering
    \scriptsize
    \renewcommand{\arraystretch}{1.06}
    \caption{
        Uncertainties and fixed assumptions used in the reactor-flux
        projections. Unless otherwise stated, normalization uncertainties are
        implemented as constrained log-normal nuisance parameters.
    }
    \label{tab:analysis_uncertainties}
    \begin{tabular}{
        >{\RaggedRight\arraybackslash}p{0.29\textwidth}
        >{\RaggedRight\arraybackslash}p{0.18\textwidth}
        >{\RaggedRight\arraybackslash}p{0.41\textwidth}
    }
        \toprule
        Quantity & Assigned relative uncertainty & Treatment \\
        \midrule

        \multicolumn{3}{l}{\textit{Statistical and signal-normalization inputs}} \\
        \cmidrule(lr){1-3}
        Poisson event counts
        & Determined by the expected yield
        & Included through the binned Poisson likelihood. \\

        ES selection efficiency
        & Finite-Monte-Carlo binomial uncertainty
        & Included as a constrained signal-normalization nuisance. \\

        Spectrum-averaged ES cross section
        & 0.5\%
        & Included as a constrained signal-normalization nuisance in the
          reactor-flux fit. \\

        \midrule
        \multicolumn{3}{l}{\textit{Background-normalization inputs}} \\
        \cmidrule(lr){1-3}

        $^{8}\mathrm{Li}$ production rate
        & 8.57\%
        & \multirow[c]{4}{0.41\textwidth}{\RaggedRight
            Combined with the finite-Monte-Carlo efficiency uncertainty and
            profiled.
          } \\
        $^{12}\mathrm{B}$ production rate
        & 8.13\%
        & \\
        $^{12}\mathrm{N}$ production rate
        & 11.0\%
        & \\
        $^{16}\mathrm{N}$ production rate
        & 12.42\%
        & \\

        \cmidrule(lr){1-3}

        {Residual IBD normalization}
        & {Tagging scenarios: 0, 50, 90, and 100\%}
        & {The residual IBD yield is treated as an additional background component. No additional physical-rate uncertainty is assigned; the finite-Monte-Carlo selection-efficiency uncertainty remains active when IBD is present.} \\

        \cmidrule(lr){1-3}

        PMT $^{40}\mathrm{K}$ activity
        & 16.15\%
        & PMT sources are excluded from the {idealized reference} 
          $N_{\mathrm{prompt}}\geq8$ scenario. For
          $N_{\mathrm{prompt}}\geq20$, the central $^{40}\mathrm{K}$ yield is
          set to zero because no simulated event survives. \\
        PMT $^{208}\mathrm{Tl}$ activity
        & 25.81\%
        & \\
        PMT $^{214}\mathrm{Bi}$ activity
        & 18.58\%
        & \\

        \cmidrule(lr){1-3}

        Radon-daughter rates
        & No physical-rate uncertainty assigned
        & The radon concentration is fixed to the adopted value;
          finite-Monte-Carlo efficiency uncertainties remain active. \\

        \midrule
        \multicolumn{3}{l}{\textit{Fixed analysis assumptions}} \\
        \cmidrule(lr){1-3}

        Number of target electrons
        & None assigned
        & Fixed to the nominal detector-model value. \\

        Live time
        & None assigned
        & Fixed to the nominal exposure of 365.25 days. \\

        Additional selection-model uncertainty
        & None assigned
        & {No additional uncertainty associated with
reconstruction performance or event-selection modeling is included in the
present projection.} \\

        Directional template shapes
        & None assigned
        & Signal and background shapes are treated as exact; no bin-by-bin
          finite-Monte-Carlo or detector-model shape nuisance is included. \\

        Cosmogenic energy-scaling exponent
        & None assigned
        & Fixed to $\alpha_{i}=0.75$. \\

        {Reactor-spectrum shape}
        & {None assigned}
        & {Fixed to the adopted reactor-spectrum model;
uncertainties associated with spectral-shape and fuel-composition
(fission-fraction) variations are not included.}
        \\

        {Fission yield and reactor-core powers}
        & {None assigned}
        & {Fixed to the nominal reactor inputs used in the
flux calculation; time-dependent core-power variations are not modeled.}
        \\

        {Reactor baselines}
        & {None assigned}
        & {Fixed to the nominal core-to-detector distances.}
        \\

        {Detector-response modeling}
        & {None assigned}
        & {Optical properties and PMT-response parameters are fixed to the reference detector simulation; associated modeling uncertainties are not included.}
        \\

        \bottomrule
    \end{tabular}
\end{table}

The quantities listed as fixed assumptions are not profiled, and their
possible uncertainties are therefore not included in the quoted
reactor-flux-normalization intervals. Only the finite-Monte-Carlo uncertainty
of the integrated selection efficiencies is propagated; statistical
fluctuations or modeling errors in individual template bins are not included.
The quoted intervals are consequently conditional on the adopted signal and
background shapes.

In particular, uncertainties associated with the reactor-spectrum shape,
fission yield, reactor-core powers and baselines, detector-response modeling,
and reconstruction and event-selection modeling are not quantified in the
present projection. These effects would need to be evaluated for a complete
experimental systematic-uncertainty budget.

\subsection{Expected event yields}
\label{subsec:expected_yields}

The expected yields are obtained from the physical interaction rates, the
simulated selection efficiencies, and the 365.25-day exposure. The unselected
ES rate is $3.43\times10^{3}$ events per day. Table~\ref{tab:all_energy_yields}
summarizes the selected all-energy yields for the two prompt-hit
requirements. These yields define the IBD-free reference and are
equivalent to the idealized 100\% IBD-tagging case used in the residual-IBD
study below.

\begin{table}[t]
    \centering
    \small
    \caption{
        Expected all-energy event yields for an exposure of 365.25 days.
        {The tabulated background yields exclude residual
        IBD and correspond to the 100\% IBD-tagging reference.}
    }
    \label{tab:all_energy_yields}
    \begin{tabular}{lcccc}
        \toprule
        Prompt-hit requirement
        & ES efficiency
        & Signal
        & Background
        & $S/B$ \\
        \midrule
        $N_{\mathrm{prompt}}\geq8$
        & 0.14361
        & 180,119.6
        & 665,269.7
        & 0.271 \\

        $N_{\mathrm{prompt}}\geq20$
        & 0.07487
        & 93,904.0
        & 583,135.4
        & 0.161 \\
        \bottomrule
    \end{tabular}
\end{table}

The more restrictive $N_{\mathrm{prompt}}\geq20$ requirement reduces the ES
selection efficiency from 0.14361 to 0.07487. Because the background
composition also changes between the two scenarios, however, the differences
in the total background yield and $S/B$ cannot be attributed to the
prompt-hit requirement alone.

\subsection{All-energy reactor-flux projection}
\label{subsec:all_energy_performance}

The nominal projections in this subsection use the IBD-free
reference background model, corresponding to idealized complete rejection of
IBD events. The impact of incomplete IBD tagging is quantified separately in
section~\ref{subsec:ibd_tagging_impact}.

\begin{figure}[t]
    \centering
    \begin{subfigure}[t]{0.49\textwidth}
        \centering
        \includegraphics[width=\linewidth]
        {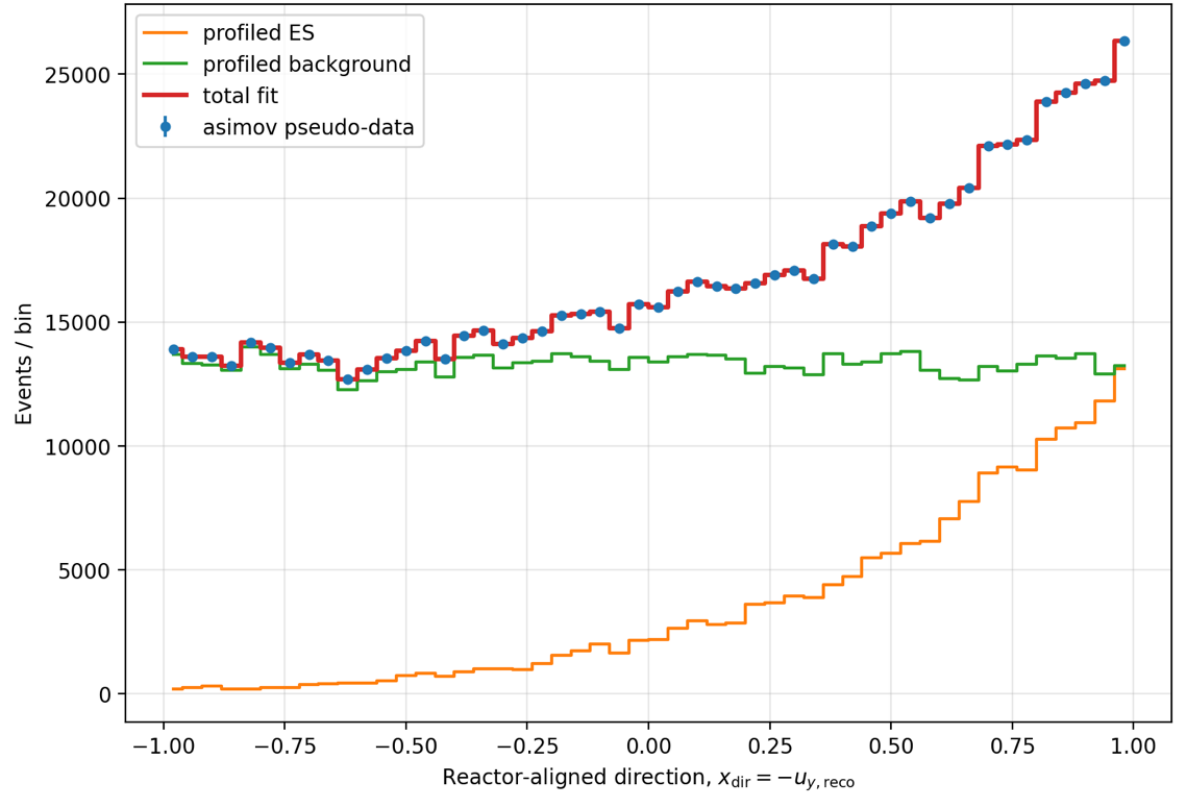}
        \caption{$N_{\mathrm{prompt}}\geq8$.}
        \label{fig:all_energy_fit_nprompt8}
    \end{subfigure}
    \hfill
    \begin{subfigure}[t]{0.49\textwidth}
        \centering
        \includegraphics[width=\linewidth]
        {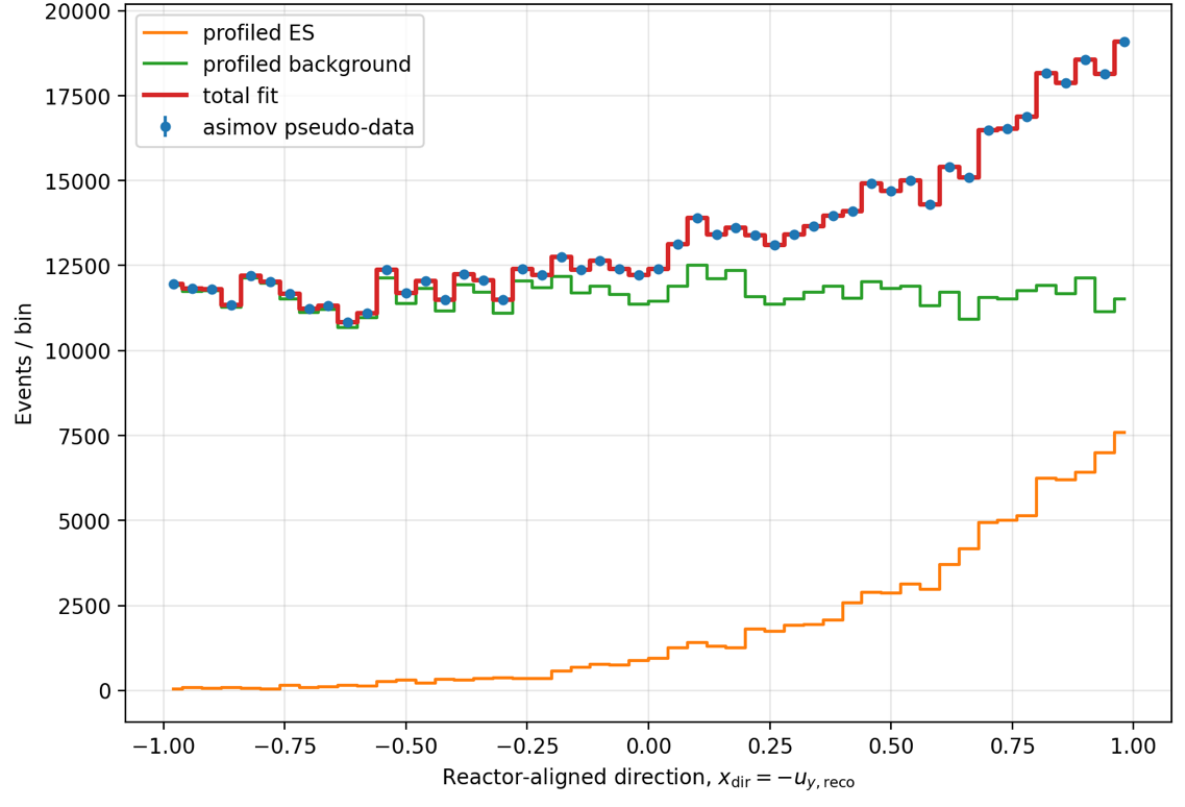}
        \caption{$N_{\mathrm{prompt}}\geq20$.}
        \label{fig:all_energy_fit_nprompt20}
    \end{subfigure}
    \caption{
        Constrained profile-likelihood fits to the all-energy Asimov samples
        for the two prompt-hit requirements. The profiled ES signal, total
        background, and total expectation are shown together with the Asimov
        bin contents. These fits correspond to the
        IBD-free reference used for the nominal all-energy projection.
    }
    \label{fig:all_energy_fits}
\end{figure}

Figure~\ref{fig:all_energy_fits} shows the constrained
profile-likelihood fits to the all-energy Asimov samples. In both scenarios,
the forward enhancement near $x_{\mathrm{dir}}=1$ distinguishes the ES
signal from the broader background distributions and provides sensitivity
to the reactor-flux normalization.

For $N_{\mathrm{prompt}}\geq8$, the projected reactor flux is
\begin{align}
    \Phi_{\bar{\nu}_e}
    ={}&
    1.449\times10^{11}
    \,\mathrm{cm^{-2}\,s^{-1}}
    \nonumber\\
    &{}^{+5.95\times10^{8}}_{-5.94\times10^{8}}
    \;\text{(stat.)}
    \nonumber\\
    &{}^{+1.56\times10^{9}}_{-1.54\times10^{9}}
    \;\text{(total)}.
    \label{eq:flux_all_nprompt8}
\end{align}
The corresponding relative statistical uncertainty is approximately
$^{+0.411\%}_{-0.410\%}$, while the total uncertainty is
$^{+1.08\%}_{-1.07\%}$.

For $N_{\mathrm{prompt}}\geq20$, the projected reactor flux is
\begin{align}
    \Phi_{\bar{\nu}_e}
    ={}&
    1.449\times10^{11}
    \,\mathrm{cm^{-2}\,s^{-1}}
    \nonumber\\
    &{}^{+9.38\times10^{8}}_{-9.37\times10^{8}}
    \;\text{(stat.)}
    \nonumber\\
    &{}^{+2.16\times10^{9}}_{-2.13\times10^{9}}
    \;\text{(total)}.
    \label{eq:flux_all_nprompt20}
\end{align}
The corresponding relative statistical uncertainty is
$^{+0.648\%}_{-0.647\%}$, while the total uncertainty is
$^{+1.49\%}_{-1.47\%}$.

The larger statistical uncertainty for
$N_{\mathrm{prompt}}\geq20$ primarily reflects the lower selected ES yield.
Because the background composition also differs, the numerical difference
between the two projections should be interpreted as a comparison of the two
analysis scenarios rather than as the effect of the prompt-hit requirement
alone.

\begingroup
\subsection{Impact of residual inverse-beta-decay background}
\label{subsec:ibd_tagging_impact}

The dedicated IBD simulation provides the prompt-event selection efficiency
before any neutron-tagging requirement. Of the $1.0\times10^{5}$ generated
IBD events, 46,428 survive the $N_{\mathrm{prompt}}\geq8$ selection and
31,171 survive the $N_{\mathrm{prompt}}\geq20$ selection, corresponding to
selection efficiencies of 0.46428 and 0.31171, respectively. With the nominal
untagged interaction rate of $1.26\times10^{4}\,\mathrm{day^{-1}}$, these
fractions correspond to approximately $2.14\times10^{6}$ and
$1.43\times10^{6}$ selected IBD events, respectively, over 365.25 days if no
neutron-tag rejection is applied.

Because the present simulation does not model neutron capture or a specific
Gd-loading configuration, the IBD rejection is treated as a parametric
sensitivity study rather than a prediction of detector performance. Four
assumed tagging efficiencies, 0\%, 50\%, 90\%, and 100\%, are considered. In
each scenario the simulated IBD directional shape is kept unchanged, and the
fraction of IBD events not removed by the assumed tag is included as an
additional background component in the same directional likelihood as the
other backgrounds. The 100\% case therefore reproduces the IBD-free nominal
projection of sections~\ref{subsec:expected_yields} and
\ref{subsec:all_energy_performance}. These assumed efficiencies should not be
interpreted as predicted Gd-tagging efficiencies for the reference detector. The resulting residual IBD yields, total background levels,
signal-to-background ratios, and projected reactor-flux uncertainties
for the four assumed tagging efficiencies are summarized in
table~\ref{tab:ibd_tagging_scenarios}.

\begin{table}[t]
    \centering
    \scriptsize
    \caption{
        Dependence of the one-year all-energy reactor-flux projection on the
        assumed IBD neutron-tagging efficiency. The residual IBD column gives
        the selected IBD events remaining after the assumed rejection. The
        quoted statistical and total uncertainties are relative 68\%
        profile-likelihood intervals.
    }
    \label{tab:ibd_tagging_scenarios}
    \resizebox{\textwidth}{!}{%
    \begin{tabular}{lrrrrcc}
        \toprule
        Prompt-hit requirement
        & Assumed IBD tagging
        & Residual IBD
        & Total background
        & $S/B$
        & Statistical $\delta\Phi/\Phi$
        & Total $\delta\Phi/\Phi$ \\
        \midrule
        $N_{\mathrm{prompt}}\geq8$
        & 0\%   & 2,136,686.2 & 2,801,955.9 & 0.0643
        & ${}^{+0.698\%}_{-0.698\%}$
        & ${}^{+1.346\%}_{-1.333\%}$ \\
        $N_{\mathrm{prompt}}\geq8$
        & 50\%  & 1,068,343.1 & 1,733,612.8 & 0.1039
        & ${}^{+0.573\%}_{-0.573\%}$
        & ${}^{+1.219\%}_{-1.207\%}$ \\
        $N_{\mathrm{prompt}}\geq8$
        & 90\%  & 213,668.6 & 878,938.3 & 0.2049
        & ${}^{+0.448\%}_{-0.448\%}$
        & ${}^{+1.107\%}_{-1.096\%}$ \\
        $N_{\mathrm{prompt}}\geq8$
        & 100\% & 0.0 & 665,269.7 & 0.2707
        & ${}^{+0.410\%}_{-0.410\%}$
        & ${}^{+1.076\%}_{-1.066\%}$ \\
        \addlinespace
        $N_{\mathrm{prompt}}\geq20$
        & 0\%   & 1,434,536.2 & 2,017,671.6 & 0.0465
        & ${}^{+1.054\%}_{-1.053\%}$
        & ${}^{+1.879\%}_{-1.856\%}$ \\
        $N_{\mathrm{prompt}}\geq20$
        & 50\%  & 717,268.1 & 1,300,403.5 & 0.0722
        & ${}^{+0.875\%}_{-0.874\%}$
        & ${}^{+1.697\%}_{-1.675\%}$ \\
        $N_{\mathrm{prompt}}\geq20$
        & 90\%  & 143,453.6 & 726,589.0 & 0.1292
        & ${}^{+0.699\%}_{-0.698\%}$
        & ${}^{+1.535\%}_{-1.514\%}$ \\
        $N_{\mathrm{prompt}}\geq20$
        & 100\% & 0.0 & 583,135.4 & 0.1610
        & ${}^{+0.647\%}_{-0.647\%}$
        & ${}^{+1.491\%}_{-1.471\%}$ \\
        \bottomrule
    \end{tabular}%
    }
\end{table}

The effect is largest when no neutron tagging is assumed. For
$N_{\mathrm{prompt}}\geq8$, the total background increases from
$6.65\times10^{5}$ in the IBD-free reference to $2.80\times10^{6}$ events,
reducing $S/B$ from 0.271 to 0.064 and increasing the total reactor-flux
uncertainty from approximately 1.07\% to 1.34\%. A 90\% assumed tagging
efficiency reduces the residual IBD sample to $2.14\times10^{5}$ events and
restores the total uncertainty to approximately 1.10\%. For
$N_{\mathrm{prompt}}\geq20$, the corresponding total uncertainty changes from
approximately 1.48\% in the IBD-free reference to 1.87\% with no tagging and
approximately 1.52\% for 90\% tagging. The fitted central value remains at
the injected value in these matched Asimov studies; residual IBD therefore
appears primarily as a degradation of statistical and profiled sensitivity
rather than a bias in the nominal fit.
\endgroup

\subsection{Idealized projections with event-level true-energy selections}
\label{subsec:true_energy_performance}

To examine how the projected sensitivity is distributed between lower- and
higher-energy events, the selected samples are divided at \SI{3}{MeV}
according to $T_{e}^{\mathrm{true}}$. As defined above,
$T_{e}^{\mathrm{true}}$ denotes the kinetic energy of the primary particle
generated in the RAT-PAC simulation. For the ES signal, this variable
corresponds to the recoil-electron kinetic energy. Independent
profile-likelihood fits are performed for each prompt-hit requirement using
the selections
\begin{equation}
    T_{e}^{\mathrm{true}}< \SI{3}{MeV}
    \qquad\text{and}\qquad
    T_{e}^{\mathrm{true}}\geq\SI{3}{MeV}.
\end{equation}
The same directional observable,
likelihood construction, and nuisance model used in the all-energy analysis
are retained.

Because $T_{e}^{\mathrm{true}}$ follows the primary-particle kinetic energy, the PMT-radioactivity samples are concentrated primarily in the
low-energy region. Their prominent gamma emissions extend to approximately
\SI{2.6}{MeV}, below the \SI{3}{MeV} boundary, and the PMT contribution is
therefore strongly suppressed in the high-energy sample. In the PMT-inclusive
$N_{\mathrm{prompt}}\geq20$ model, the residual high-energy contribution is
subdominant and is retained explicitly. In contrast, the beta and gamma spectra
associated with the radon-daughter and cosmogenic-isotope backgrounds extend
to several MeV and above. Substantial fractions of these components
consequently remain in the high-energy sample. The energy selection therefore
changes the relative composition of the background rather than suppressing all
background sources uniformly.

Table~\ref{tab:true_energy_flux_results} summarizes the expected event yields,
signal-to-background ratios, and projected reactor-flux uncertainties for an
exposure of 365.25 days. The comparison focuses on the relative statistical
and total uncertainties obtained for each event-level true-energy selection.

\begin{table}[t]
    \centering
    \scriptsize
    \caption{
        Expected one-year event yields and projected reactor-flux uncertainties
        for the idealized event-level true-energy selections. The quoted statistical
        and total uncertainties are relative 68\% profile-likelihood intervals.
        The PMT-radioactivity treatment is indicated explicitly for each sample.
    }
    \label{tab:true_energy_flux_results}
    \resizebox{\textwidth}{!}{%
    \begin{tabular}{llcrrrcc}
        \toprule
        Prompt-hit requirement
        & True-energy selection
        & PMT radioactivity
        & $N_{\mathrm{ES}}$
        & $N_{\mathrm{bkg}}$
        & $S/B$
        & Statistical $\delta\Phi/\Phi$
        & Total $\delta\Phi/\Phi$ \\
        \midrule
        $N_{\mathrm{prompt}}\geq8$
        & $T_{e}^{\mathrm{true}}<\SI{3}{MeV}$
        & Excluded
        & 136,434.8
        & 139,049.2
        & 0.981
        & $^{+0.350\%}_{-0.350\%}$
        & $^{+1.133\%}_{-1.120\%}$ \\

        $N_{\mathrm{prompt}}\geq8$
        & $T_{e}^{\mathrm{true}}\geq\SI{3}{MeV}$
        & Subdominant
        & 43,684.7
        & 526,220.5
        & 0.083
        & $^{+1.177\%}_{-1.175\%}$
        & $^{+2.304\%}_{-2.260\%}$ \\

        $N_{\mathrm{prompt}}\geq20$
        & $T_{e}^{\mathrm{true}}<\SI{3}{MeV}$
        & Included
        & 52,577.2
        & 43,962.7
        & 1.196
        & $^{+0.538\%}_{-0.537\%}$
        & $^{+1.732\%}_{-1.703\%}$ \\

        $N_{\mathrm{prompt}}\geq20$
        & $T_{e}^{\mathrm{true}}\geq\SI{3}{MeV}$
        & \shortstack{Subdominant}
        & 41,326.8
        & 539,172.7
        & 0.077
        & $^{+1.249\%}_{-1.247\%}$
        & $^{+2.399\%}_{-2.353\%}$ \\
        \bottomrule
    \end{tabular}%
    }
\end{table}

For both prompt-hit requirements, the low-energy truth subset yields a smaller
projected normalization uncertainty than the corresponding high-energy subset.
For $N_{\mathrm{prompt}}\geq8$, the low-energy sample retains approximately
$1.36\times10^{5}$ ES events while reducing the selected non-PMT background to
a comparable level. This gives $S/B\simeq0.98$ and a total normalization
uncertainty of approximately 1.1\%. The high-energy sample contains fewer ES
events and remains strongly background dominated because substantial
radon-daughter and cosmogenic-isotope contributions survive the energy
selection, giving a total uncertainty of approximately 2.3\%.

The same qualitative behavior is observed for
$N_{\mathrm{prompt}}\geq20$. With the corrected component-by-component
normalization, the low-energy selection contains approximately
$4.40\times10^{4}$ background events, gives $S/B\simeq1.20$, and has a total
uncertainty of approximately 1.7\%. The high-energy selection gives
$S/B\simeq0.077$ and a total uncertainty of approximately 2.4\%. The residual
PMT contribution is subdominant in this high-energy sample, while the
radon-daughter and cosmogenic-isotope components dominate.

The different responses of the individual background sources to the
truth-level recoil-energy selection illustrate the potential role of energy
information in future analysis optimization. In particular, the
$N_{\mathrm{prompt}}\geq8$ result below 3~MeV suggests that retaining
lower-light events could improve the reactor-flux precision if PMT
radioactivity can be sufficiently controlled and a suitable recoil-energy
reconstruction can be achieved.

\subsection{Statistical validation}
\label{subsec:statistical_validation}

The likelihood procedure is validated with ensembles of pseudo-experiments
for the all-energy $N_{\mathrm{prompt}}\geq20$ scenario. The pseudo-data are
generated with $\alpha_{\mathrm{true}}=1$ using the same signal and background
templates as the fit, including the PMT-radioactivity components. This
matched-model test evaluates estimator bias, pull behavior, and interval
coverage under the adopted model.

Two ensembles, each containing 5,000 pseudo-experiments, are considered. In
the \emph{statistical-only} ensemble, the binned event counts are independently
Poisson fluctuated, while all nuisance parameters are fixed at their injected
values. The signal strength is then fitted using the conditional statistical
interval defined by $2\Delta\mathrm{NLL}=1$.

In the \emph{total-auxiliary} ensemble, both the binned event counts and the
auxiliary measurements associated with the constrained nuisance parameters
are independently fluctuated in each pseudo-experiment. The signal strength
and all active nuisance parameters are then profiled simultaneously. This
ensemble tests the frequentist coverage of the total profile-likelihood
interval, including the effects of the normalization uncertainties represented
by the nuisance parameters.

Because the profile-likelihood intervals can be asymmetric, the pull is
defined using the uncertainty on the side of the fitted value that contains
the injected truth,
\begin{equation}
    p
    =
    \frac{\hat{\alpha}-\alpha_{\mathrm{true}}}
         {\sigma_{\mathrm{toward\ truth}}}.
    \label{eq:pull_definition}
\end{equation}
Here, $\sigma_{\mathrm{toward\ truth}}$ is the lower uncertainty when
$\hat{\alpha}>\alpha_{\mathrm{true}}$ and the upper uncertainty when
$\hat{\alpha}<\alpha_{\mathrm{true}}$. An unbiased and statistically
consistent fit is expected to yield
$\langle\hat{\alpha}\rangle\simeq1$, a pull mean near zero, a pull width near
unity, and a 68\% interval coverage consistent with the nominal value of
68.27\%.

The results are summarized in table~\ref{tab:pull_validation}. In both
ensembles, the fitted signal strength is consistent with the injected value,
the pull mean is consistent with zero, the pull width is consistent with
unity, and the 68\% coverage agrees with the nominal expectation. The largest
reported relative bias is 0.0091\%.

\begin{figure}[t]
    \centering
    \begin{subfigure}[t]{0.49\textwidth}
        \centering
        \includegraphics[width=\linewidth]
        {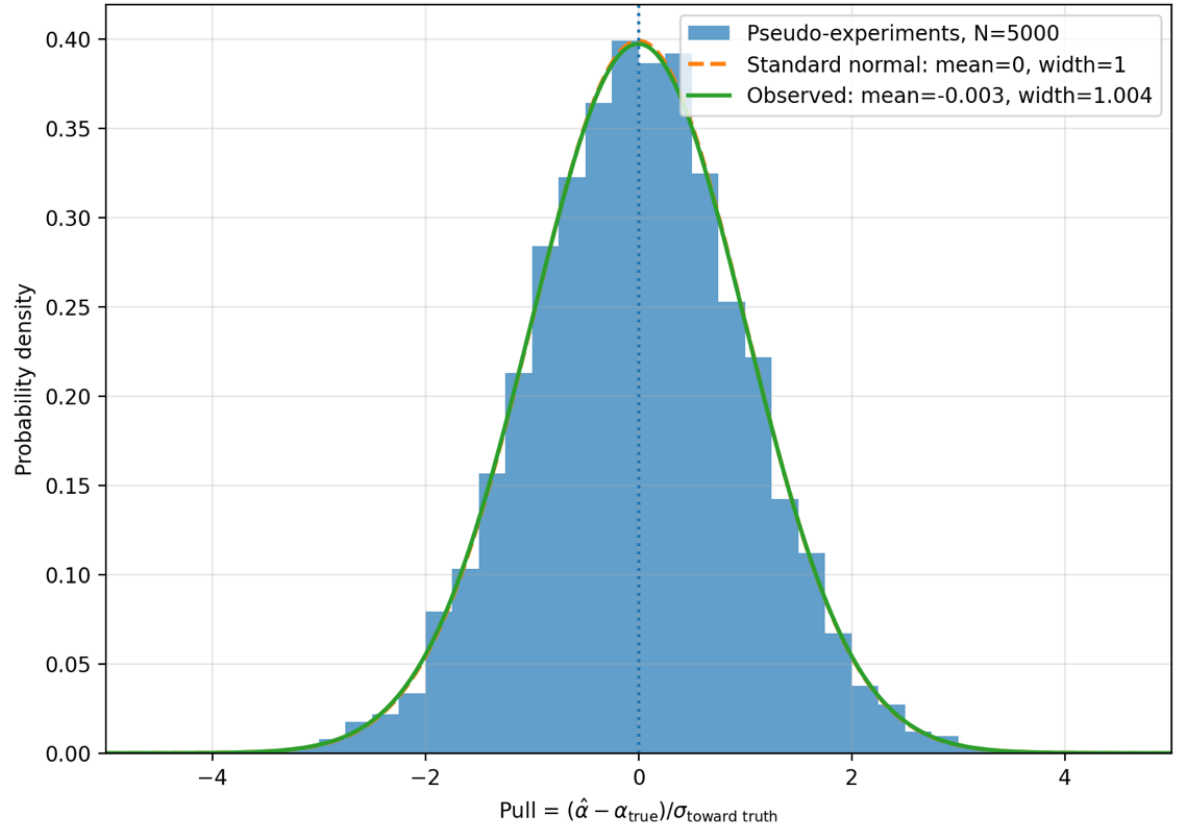}
        \caption{
            Pull distribution for the total profile-likelihood interval.
        }
        \label{fig:pull_distribution_total}
    \end{subfigure}
    \hfill
    \begin{subfigure}[t]{0.49\textwidth}
        \centering
        \includegraphics[width=\linewidth]
        {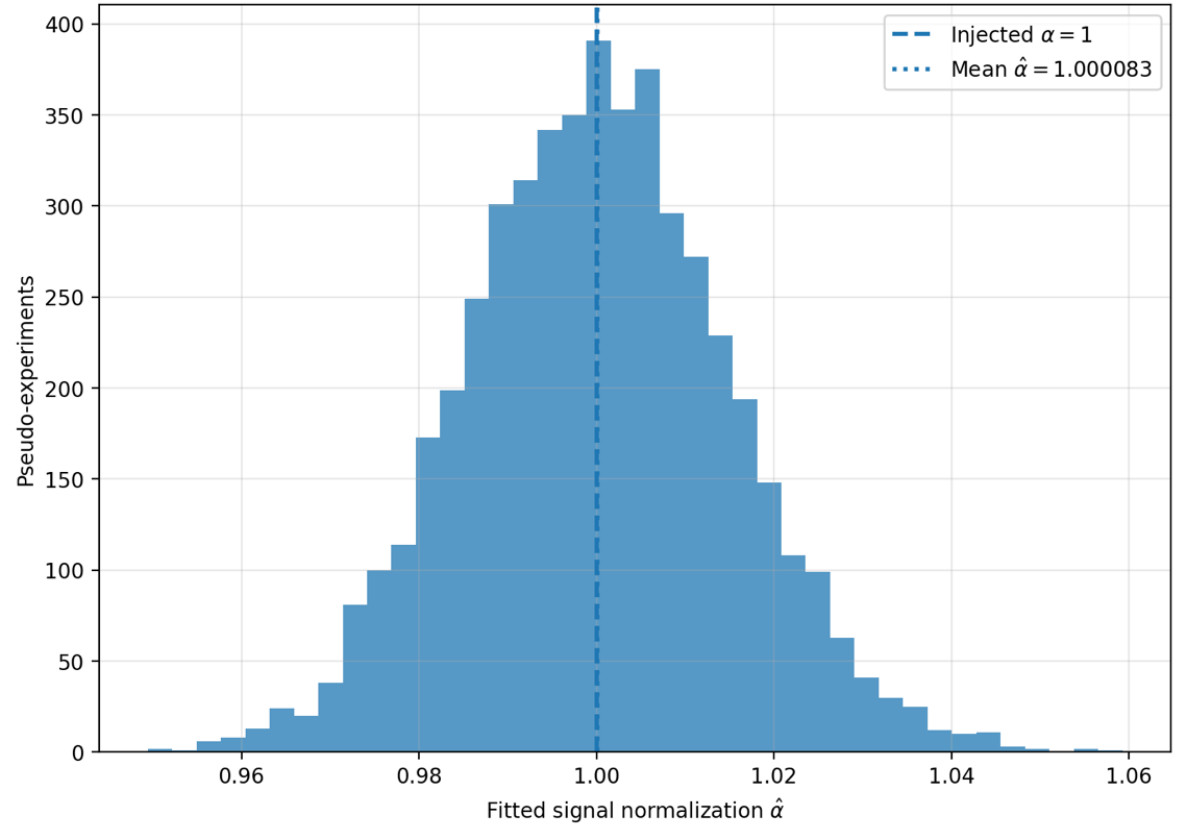}
        \caption{
            Distribution of the fitted ES signal strength
            $\hat{\alpha}$.
        }
        \label{fig:alpha_distribution_total}
    \end{subfigure}
    \caption{
        Statistical validation of the matched all-energy model with
        $N_{\mathrm{prompt}}\geq20$ using 5,000 total-auxiliary
        pseudo-experiments. The dashed curve in the left panel represents a
        standard normal distribution. In the right panel, the dashed vertical
        line indicates the injected signal strength
        $\alpha_{\mathrm{true}}=1$.
    }
    \label{fig:pull_validation_nprompt20}
\end{figure}

Figure~\ref{fig:pull_validation_nprompt20} shows the pull and fitted
signal-strength distributions for the total-auxiliary ensemble. Their
agreement with the expected standard-normal pull and the injected signal
strength demonstrates closure of the likelihood implementation and validates the statistical behavior of the profile-likelihood intervals under the adopted model.

\begin{table}[t]
    \centering
    \small
    \caption{
        Pull-test results for the all-energy configuration with
        $N_{\mathrm{prompt}}\geq20$. Each ensemble contains 5,000
        pseudo-experiments generated with an injected signal strength
        $\alpha_{\mathrm{true}}=1$.
    }
    \label{tab:pull_validation}
    \begin{tabular}{lcccc}
        \toprule
        Ensemble
        & $\langle\hat{\alpha}\rangle$
        & Pull mean
        & Pull width
        & 68\% coverage \\
        \midrule
        Statistical only
        & $1.000091\pm0.000092$
        & $0.0135\pm0.0142$
        & $1.0014\pm0.0100$
        & $(68.96\pm0.65)$\% \\
        Total auxiliary
        & $1.000083\pm0.000210$
        & $-0.0026\pm0.0142$
        & $1.0042\pm0.0100$
        & $(68.14\pm0.66)$\% \\
        \bottomrule
    \end{tabular}
\end{table}

\clearpage
\section{Conclusion}
\label{sec:conclusion}

The principal physics result is obtained from the all-energy directional fit
without event-by-event recoil-energy reconstruction. The nominal
results quoted first correspond to the IBD-free reference, equivalent to
idealized complete rejection of IBD events. For an exposure of
365.25 days, the analysis assumes the adopted reactor-spectrum shape
and a spectrum-averaged ES cross section of
$4.821 \times 10^{-45}\,\mathrm{cm}^{2}$ with a theoretical uncertainty of
0.5\%. The $N_{\mathrm{prompt}} \geq 8$ selection yields a total uncertainty of
${}^{+1.08\%}_{-1.07\%}$ on the nominal reactor-spectrum normalization.
The more restrictive $N_{\mathrm{prompt}} \geq 20$ selection,
which includes the simulated PMT-radioactivity sources, yields a
corresponding total uncertainty of ${}^{+1.49\%}_{-1.47\%}$.
The quoted reactor-spectrum normalization is model dependent, as it is
determined within the adopted reactor-spectrum model with fixed spectral shape,
fission yield, core powers, and baselines, and subject to the external
cross-section constraint. Because the fitted ES normalization constrains the
product of the reactor-spectrum normalization and the spectrum-averaged cross
section, the result can also be interpreted as a cross-section measurement
when an external constraint on the reactor-spectrum normalization is available.

A dedicated prompt-IBD background study shows that the IBD contribution cannot
be assumed negligible at the RENO near-detector baselines if the neutron is not
identified. For $N_{\mathrm{prompt}}\geq8$ and
$N_{\mathrm{prompt}}\geq20$, the total reactor-flux uncertainties in the
no-tagging scenario are ${}^{+1.35\%}_{-1.33\%}$ and
${}^{+1.88\%}_{-1.86\%}$, respectively. Under an assumed 90\%
IBD-tagging efficiency, the total uncertainties improve to
${}^{+1.11\%}_{-1.10\%}$ and ${}^{+1.53\%}_{-1.51\%}$, respectively,
approaching the IBD-free reference values. These tagging efficiencies are
parametric assumptions: the present simulation includes the prompt positron
response but does not simulate neutron thermalization, capture, or a specific
Gd-loaded detector response.

Idealized selections based on the Monte Carlo primary-particle kinetic energy
were also studied to assess the sensitivity in different energy regions. For
the $N_{\mathrm{prompt}} \geq 8$ selection, the total reactor-spectrum
normalization uncertainties are approximately 1.1\% and 2.3\% for
$T_{e}^{\mathrm{true}} < \SI{3}{MeV}$ and
$T_{e}^{\mathrm{true}} \geq \SI{3}{MeV}$, respectively. For the
$N_{\mathrm{prompt}} \geq 20$ selection, approximately
$5.26 \times 10^{4}$ ES events are retained below \SI{3}{MeV}, giving a total
uncertainty of approximately 1.7\%, while the corresponding high-energy sample
gives a total uncertainty of approximately 2.4\%. These truth-level results
require a validated event-by-event recoil-energy reconstruction for
experimental application. 
Accordingly, the 3~MeV-separated results should be interpreted as
idealized truth-level projections rather than as directly achievable
experimental sensitivities. Their realization would depend on the resolution,
bias, and selection performance of a future recoil-energy reconstruction.

Additional
quantitative studies are required to evaluate uncertainties associated with
the number of target electrons, the live time, detector-response modeling,
event-selection modeling, and the cosmogenic energy-scaling exponent. A
realistic experimental implementation will also require data-driven,
site-specific background templates, direct measurements of radioactivity in
the PMTs and detector materials, continuous monitoring of radon levels and
water quality, and muon-tagged control samples for cosmogenic backgrounds.
A realistic assessment of IBD rejection will additionally require
explicit neutron transport and capture simulation, a specified Gd loading,
and validation of the delayed-coincidence tagging efficiency and accidental
background rate.
These measurements should be incorporated into the optimization of the
detector geometry, photosensor configuration, veto design, and analysis
thresholds. A quantitative assessment of these experimental and modeling
effects will be pursued in future work.

Within the assumptions adopted in this study, the results demonstrate that
directional information from a water Cherenkov detector can provide
percent-level sensitivity to the reactor ES normalization. They also indicate
that validated recoil-energy reconstruction and improved background
characterization are the principal requirements for extending the physics
reach of this approach.

\appendix
\appendix

\section{Water optical properties}
\label{app:water_optical_properties}

The wavelength-dependent optical properties of water used in the RAT-PAC
simulation are summarized in table~\ref{tab:water_optical_properties}. The
refractive index was calculated using the IAPWS formulation for water at
\SI{295}{K} and a density of \SI{1000}{kg.m^{-3}}, while the absorption and
Rayleigh-scattering lengths were obtained from
refs.~\cite{FewellTrojan2019,ZhangHu2021}, respectively.

\begin{longtable}{cccc}
    \caption{
        Representative wavelength-dependent optical properties of water used
        in the RAT-PAC simulation. The absorption and Rayleigh-scattering
        lengths are given in metres.
    }
    \label{tab:water_optical_properties} \\

    \toprule
    Wavelength [nm]
    & Refractive index
    & $L_{\mathrm{abs}}$ [m]
    & $L_{\mathrm{Ray}}$ [m] \\
    \midrule
    \endfirsthead

    \multicolumn{4}{c}{
        \tablename\ \thetable\ -- continued from previous page
    } \\
    \toprule
    Wavelength [nm]
    & Refractive index
    & $L_{\mathrm{abs}}$ [m]
    & $L_{\mathrm{Ray}}$ [m] \\
    \midrule
    \endhead

    \midrule
    \multicolumn{4}{r}{Continued on next page} \\
    \endfoot

    \bottomrule
    \endlastfoot

    200   & 1.4250 & 1.661  & 6.813  \\
    205   & 1.4174 & 4.274  & 7.922  \\
    210   & 1.4108 & 6.250  & 9.132  \\
    215   & 1.4050 & 7.299  & 10.45  \\
    220   & 1.3999 & 9.901  & 11.87  \\
    225   & 1.3953 & 13.33  & 13.42  \\
    230   & 1.3912 & 14.29  & 15.09  \\
    235   & 1.3875 & 15.38  & 16.88  \\
    240   & 1.3842 & 17.54  & 18.81  \\
    245   & 1.3812 & 23.81  & 20.89  \\
    250   & 1.3784 & 16.95  & 23.11  \\
    260   & 1.3735 & 19.23  & 28.02  \\
    270   & 1.3693 & 22.73  & 33.61  \\
    280   & 1.3657 & 44.84  & 39.93  \\
    290   & 1.3626 & 106.4  & 47.03  \\
    300   & 1.3598 & 212.8  & 54.98  \\
    310   & 1.3574 & 416.7  & 63.84  \\
    320   & 1.3553 & 666.7  & 73.68  \\
    330   & 1.3534 & 909.1  & 84.56  \\
    340   & 1.3516 & 1176.5 & 96.55  \\
    350   & 1.3501 & 1123.6 & 109.7  \\
    360   & 1.3487 & 869.6  & 124.1  \\
    370   & 1.3474 & 806.5  & 139.9  \\
    380   & 1.3462 & 699.3  & 157.1  \\
    390   & 1.3451 & 588.2  & 175.7  \\
    400   & 1.3441 & 450.5  & 195.9  \\
    410   & 1.3432 & 375.9  & 217.8  \\
    420   & 1.3424 & 320.5  & 241.5  \\
    430   & 1.3416 & 266.0  & 266.9  \\
    440   & 1.3408 & 192.3  & 294.3  \\
    450   & 1.3402 & 123.5  & 323.7  \\
    460   & 1.3395 & 109.9  & 355.2  \\
    470   & 1.3389 & 97.09  & 388.9  \\
    480   & 1.3384 & 82.64  & 425.0  \\
    490   & 1.3378 & 68.49  & 463.4  \\
    500   & 1.3373 & 48.31  & 504.4  \\
    510   & 1.3369 & 30.30  & 548.0  \\
    520   & 1.3364 & 25.51  & 594.3  \\
    530   & 1.3360 & 23.58  & 643.5  \\
    540   & 1.3356 & 21.05  & 695.6  \\
    550   & 1.3352 & 17.76  & 750.8  \\
    560   & 1.3349 & 16.16  & 809.3  \\
    570   & 1.3345 & 14.41  & 871.0  \\
    580   & 1.3342 & 11.16  & 936.2  \\
    590   & 1.3339 & 7.386  & 1005.0 \\
    600   & 1.3336 & 4.480  & 1077.5 \\
    610   & 1.3333 & 3.769  & 1153.8 \\
    620   & 1.3330 & 3.623  & 1234.0 \\
    630   & 1.3327 & 3.428  & 1318.4 \\
    640   & 1.3325 & 3.220  & 1407.1 \\
    650   & 1.3322 & 2.941  & 1500.1 \\
    660   & 1.3320 & 2.439  & 1597.6 \\
    670   & 1.3317 & 2.278  & 1699.9 \\
    680   & 1.3315 & 2.151  & 1807.0 \\
    690   & 1.3313 & 1.938  & 1919.0 \\
    700   & 1.3311 & 1.603  & 2036.2 \\
    710   & 1.3308 & 1.208  & 2158.7 \\
    720   & 1.3306 & 0.809  & 2286.7 \\
    725   & 1.3305 & 0.668  & 2352.7 \\
    729.9 & 1.3304 & 0.508  & 2418.9 \\
    739.6 & 1.3302 & 0.358  & 2553.9 \\
    759.9 & 1.3299 & 0.348  & 2854.8 \\
    770.4 & 1.3297 & 0.355  & 3020.6 \\
    780   & 1.3295 & 0.372  & 3178.4 \\
    789.9 & 1.3293 & 0.407  & 3347.5 \\
    800   & 1.3291 & 0.446  & 3527.0 \\

\end{longtable}

\acknowledgments
This research was supported by the Chung-Ang University Research Scholarship Grants in 2025. This work was supported by the National Research Foundation of Korea(NRF) grant NRF-2022R1A2C1009686 and RS-2025-25460815 funded by the Korea government(MSIT).



\providecommand{\href}[2]{#2}\begingroup\raggedright\endgroup

\end{document}